\documentclass[preprint]{iucr}              % DO NOT DELETE THIS LINE
\RequirePackage{graphicx}
\usepackage{amsmath}

     \journalcode{S}              % Indicate the journal to which submitted
\begin{document}                  % DO NOT DELETE THIS LINE

     %-------------------------------------------------------------------------
     % The introductory (header) part of the paper
     %-------------------------------------------------------------------------

     % The title of the paper. Use \shorttitle to indicate an abbreviated title
     % for use in running heads (you will need to uncomment it).

\title{Very high brightness and power LCLS-II hard X-ray pulses}
%\shorttitle{Short Title}

     % Authors' names and addresses. Use \cauthor for the main (contact) author.
     % Use \author for all other authors. Use \aff for authors' affiliations.
     % Use lower-case letters in square brackets to link authors to their
     % affiliations; if there is only one affiliation address, remove the [a].
     
\cauthor[a,b]{Aliaksei}{Halavanau}{aliaksei@slac.stanford.edu}{address if different from \aff}
\author[a]{Franz-Josef}{Decker}
\author[a]{Claudio}{Emma}
%\author[a]{Yiping}{Feng}
\author[a]{Jackson}{Sheppard}
\author[a]{Claudio}{Pellegrini}

\aff[a]{SLAC National Accelerator Laboratory, Menlo Park, CA, \country{USA}}
\aff[b]{University of California, Los Angeles, CA, \country{USA} }

     % Use \shortauthor to indicate an abbreviated author list for use in
     % running heads (you will need to uncomment it).

%\shortauthor{Soape, Author and Doe}

     % Use \vita if required to give biographical details (for authors of
     % invited review papers only). Uncomment it.

%\vita{Author's biography}

     % Keywords (required for Journal of Synchrotron Radiation only)
     % Use the \keyword macro for each word or phrase, e.g. 
     % \keyword{X-ray diffraction}\keyword{muscle}

%\keyword{keyword}

     % PDB and NDB reference codes for structures referenced in the article and
     % deposited with the Protein Data Bank and Nucleic Acids Database (Acta
     % Crystallographica Section D). Repeat for each separate structure e.g
     % \PDBref[dethiobiotin synthetase]{1byi} \NDBref[d(G$_4$CGC$_4$)]{ad0002}

%\PDBref[optional name]{refcode}
%\NDBref[optional name]{refcode}

\maketitle                        % DO NOT DELETE THIS LINE

\begin{synopsis}
In this paper, we report on high peak power and brightness hard X-ray generation studies achievable in the double-bunch LCLS-II linac operation.
\end{synopsis}

\begin{abstract}
We show the feasibility of generating X-ray pulses in the 4 to 8 keV fundamental photon energy range with 0.65 TW peak power, 15 fs pulse duration and $9\times10^{-5}$ bandwidth, using the LCLS-II copper linac and hard X-ray (HXR) undulator. In addition, we generate third harmonic pulses with 8-12 GW peak power and narrow bandwidth are also generated. High power and small bandwidth X-rays are obtained using two electron bunches separated by about 1 ns, one to generate a high power seed signal, the other to amplify it through the process of the HXR undulator tapering. The bunch delay is compensated by delaying the seed pulse with a four crystal monochromator. The high power seed leads to higher output power and better spectral properties, with more than 94\% of the X-ray power within the near transform limited bandwidth. We then discuss some of the experiments made possible by X-ray pulses with these characteristics such as single particle imaging and high field physics. 
\end{abstract}

     %-------------------------------------------------------------------------
     % The main body of the paper
     %-------------------------------------------------------------------------
     % Now enter the text of the document in multiple \section's, \subsection's
     % and \subsubsection's as required.

\section{Introduction}
In this paper, we consider using the LCLS-II copper linac and the variable gap HXR undulator to implement the double bunch FEL (DBFEL) concept \cite{PhysRevSTAB.13.060703, Geloni:2010db, EmmaC}. 
In the paper \cite{EmmaC}, the DBFEL was mainly studied for the
generation of high power harmonics. In this paper we study its use to increase the
fundamental peak power and X-ray brightness. 
We also compare the results with those of other self-seeding methods, such as single bunch \cite{Amann} and fresh-slice self-seeding \cite{Lutman2016, Emma_slice, Lutman:2018cot}, showing that the DBFEL gives the highest peak power and brightness at LCLS-II.

The DBFEL is equivalent to having two FELs, the first to generate a high power, small bandwidth, seeding signal and the second to amplify it. The main advantage with respect to other LCLS self-seeding schemes using a single bunch, is to have large seed
power and pulse energy within a small bandwidth, leading, as we will show in section \ref{simulation-results}, to larger output power
and better spectral properties, and thus a large improvement in the X-ray peak brightness of LCLS-II.
Comparing to fresh slice self-seeding, DBFEL has the advantage of using for the same pulse duration electron bunches with a smaller charge and hence a smaller emittance.

This concept can be implemented in LCLS-II using two bunches from the copper linac, separated in time by about one nanosecond, and a four crystal monochromator to
delay the seed pulse by the same amount of time. Using this scheme the seed signal for the amplifier is an order of magnitude or more larger than in other single bunch self-seeding systems, an important advantage leading to increased output power and improved longitudinal coherence, as we will show in this paper. The HXR variable gap undulator allows strong tapering and high efficiency of energy transfer from the electron beam to the radiation field. The acceleration of multiple bunches in the linac, with variable time separation, needed for the double bunch system has already been achieved on the SLAC copper linac \cite{FJDecker}. 

In this paper, we first show the results of time-dependent GENESIS \cite{Genesis} simulations of the double bunch FEL using standard LCLS operating electron beam parameters. We compare, for the same beam parameters, the X-ray pulse characteristics for the proposed DBFEL with those of the single bunch, single crystal self-seeding system presently in operation \cite{Amann}. 
The paper is organized as follows.
In section \ref{section-dbfel} we consider in detail the DBFEL system and the generation of the seed signal. In section \ref{section-taper} we discuss the amplifier section tapering
strategy, in section \ref{section-mono} the monochromator design and properties, and in section \ref{section-double} the system
to generate the two bunches and control their relative timing and energy. In section \ref{simulation-results} we
present and discuss our main results on the radiation generated in the range of 4 to 8 keV. In section \ref{fresh-slice} we provide a quantitative comparison of DBFEL with the existing fresh slice technique, based on experimental LCLS performance. Finally, in section \ref{section-applications} we discuss some of the applications made possible by the availability of near TW X-ray pulses such as single particle imaging \cite{Aquila,Zhibin,Geloni:2012nb}. We also consider the possibility of focusing the photons to a spot size of 10 nm, smaller than the present value of 100 nm, for high field science. We notice that the power density obtained with a 10 nm spot size is about $2\times10^{23}$ W/cm$^2$, and corresponding peak electric field value is 10$^{15}$ V/m.

This value is larger or comparable with that obtainable with PW lasers becoming available in a few laboratories. 
Thus a DBFEL would open the possibility of exploring high field science
in the X-ray wavelength region, complimentary to what PW lasers can do in the micrometer wavelength region.

\section{High power double bunch FEL}
\label{section-dbfel}
\begin{figure}
\label{double-bunch}
\caption{Schematic of a double bunch LCLS-II undulator operation.}\includegraphics[width=0.87\linewidth]{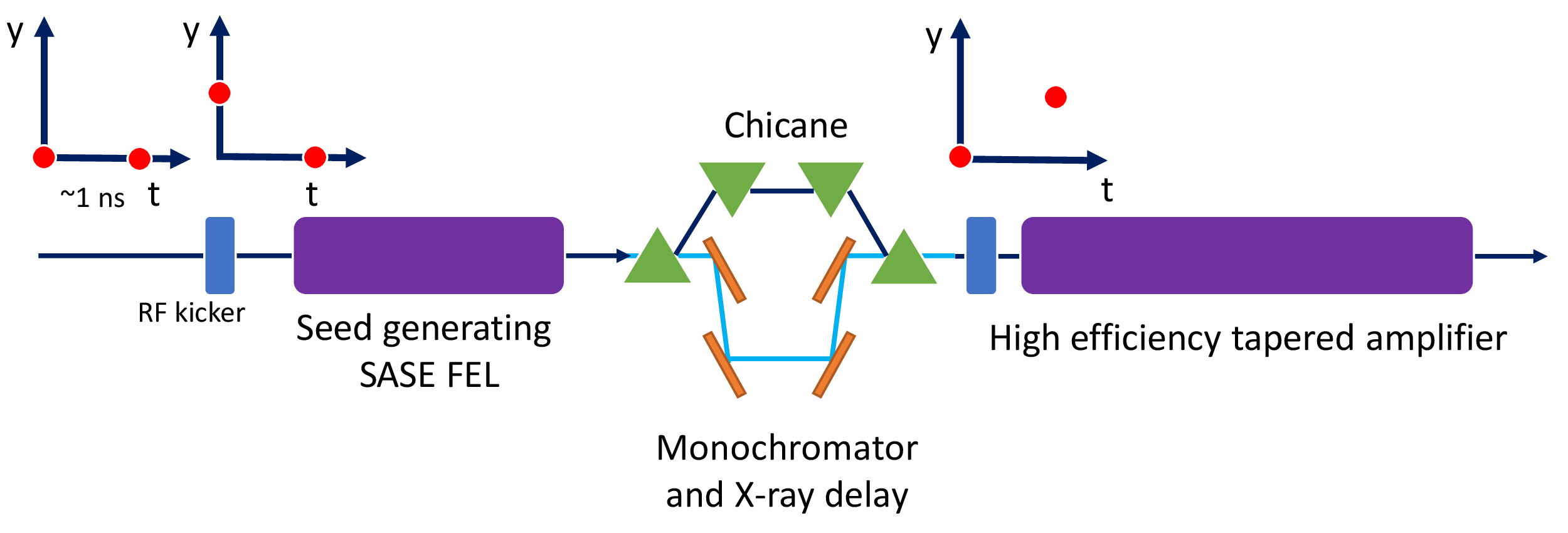}
\end{figure}

The schematic of a double bunch FEL is shown in Fig. \ref{double-bunch}.  In the first undulator section we let the first bunch lase, generating a large power, possibly reaching saturation. In the process the bunch energy spread grows to the order of the FEL parameter, about 10$^{-3}$, precluding its use in the amplifier section.  The second bunch goes through the first undulator section with a large oscillation around the axis, produced by a transverse electric field cavity, and does not lase, accumulating negligible increase in its energy spread \cite{PhysRevAccelBeams.20.040703}. At the exit of the first undulator section the first bunch is kicked out, the second bunch receives a counter kick to move on axis in the following undulator.
The radiation field is filtered through a monochromator and delayed by a time equal to the separation between the two bunches. The chicane is then used for the electron beam to bypass the monochromator crystals. At the
entrance of the tapered undulator section, the second bunch is
seeded and amplified.

The two undulators, soft X-ray (SXR) and hard X-ray (HXR), will be available for LCLS-II and are shown in Figure 2. Their main properties are given in \cite{Nuhn,osti_1029479,Lauer:2018ukq} and summarized in Tabs. \ref{lcls-beam-parameters}, \ref{lcls-undulator-parameters}.
We consider only the HXR undulator, with 32 sections, undulator period 2.6 cm, section length 340 cm, variable gap with an undulator parameter in the range of 2.4 to less than 1. The gap height can be adjusted
longitudinally, giving a magnetic field change of up to 1\% from the entrance to the exit and allowing for a smooth tapering profile \cite{Nuhn}. The separation between undulator sections is 60 cm, and two sections, 24 and 32, have a chicane for the electron beam and can be used to insert a single crystal or a multiple crystals monochromator. 
To minimize changes in the LCLS-II layout, we assume a four crystal monochromator to be placed in section 24, use the first seven sections to generate the seed signal in a SASE mode and the remaining sections, U25 to U50, to amplify the seed. The general characteristics of the copper linac, based on the operational experience of LCLS, are given in Tab. \ref{lcls-beam-parameters}. 

\begin{figure}
\label{lcls2}
\caption{Schematic of LCLS-II variable gap undulators. We propose to use the hard X-ray (HXR) undulator in the double bunch configuration, inserting a four crystal monochromator in section U24 \cite{Nuhn}.}\includegraphics[width=1\linewidth]{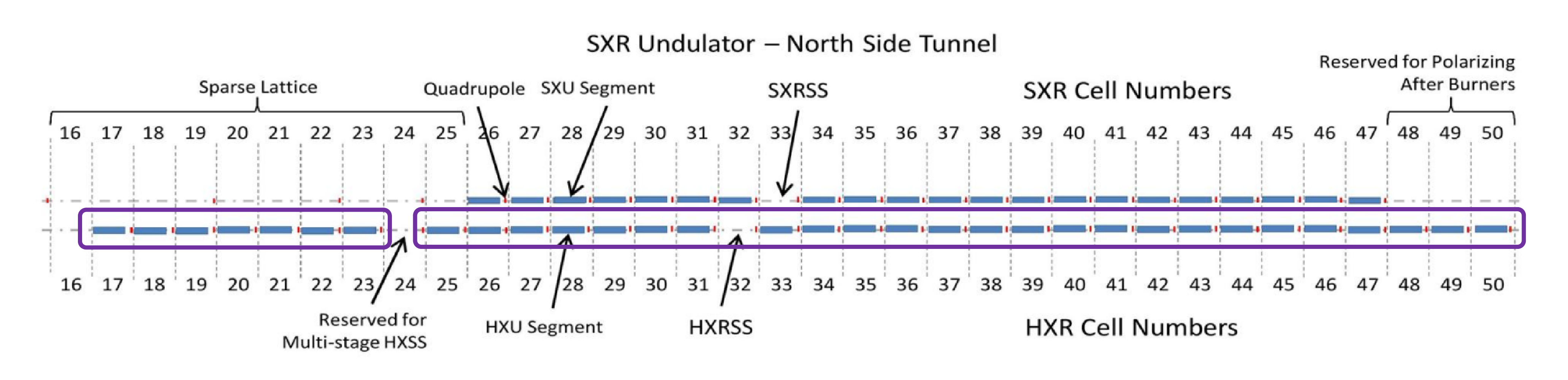}
\end{figure}

We assume that the linac generates a flat current profile bunch \cite{PhysRevAccelBeams.19.100703} to optimize the FEL performance. This is done by starting with a larger charge and
bunch length and cutting its central part with collimators in the linac bunch compressor. In the Tab. \ref{beam-parameters} case one starts with a 80 pC charge, reduced
to 60 pC after collimation. The emittance, which depends on the charge, is evaluated at 80 pC to be 0.35 $\mu$m \cite{Ding:2009zzc} and we
increase this value to 0.4 $\mu$m in Tab. \ref{beam-parameters}, to be on the conservative side. Notice that for a later comparison with the double slice FEL we use a bunch charge of 180 pC,
corresponding to an initial charge of 240 pC and a normalized emittance of 0.6 $\mu$m.

\begin{figure}
\label{Figure3}
\caption{Power gain length as a function of photon energy for two different electron beam energies, in MeV, and for an undulator parameter varying between 1 and 2.4.}\includegraphics[width=0.8\linewidth]{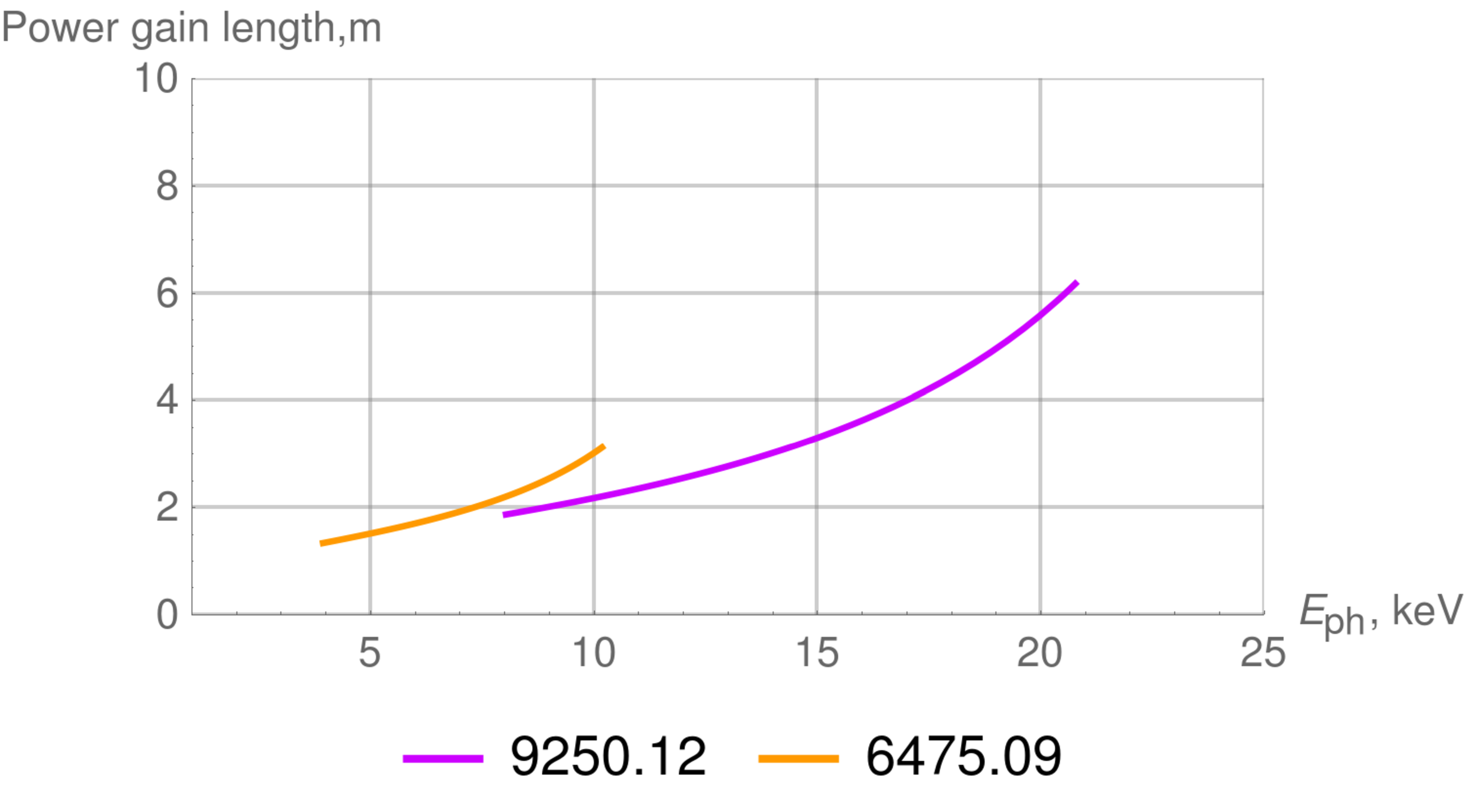}
\end{figure}

\begin{figure}
\label{Figure4}
\caption{Saturation power as a function of photon energy for two different electron beam energies, in MeV, and for an undulator parameter varying between 1 and 2.4.}\includegraphics[width=0.74\linewidth]{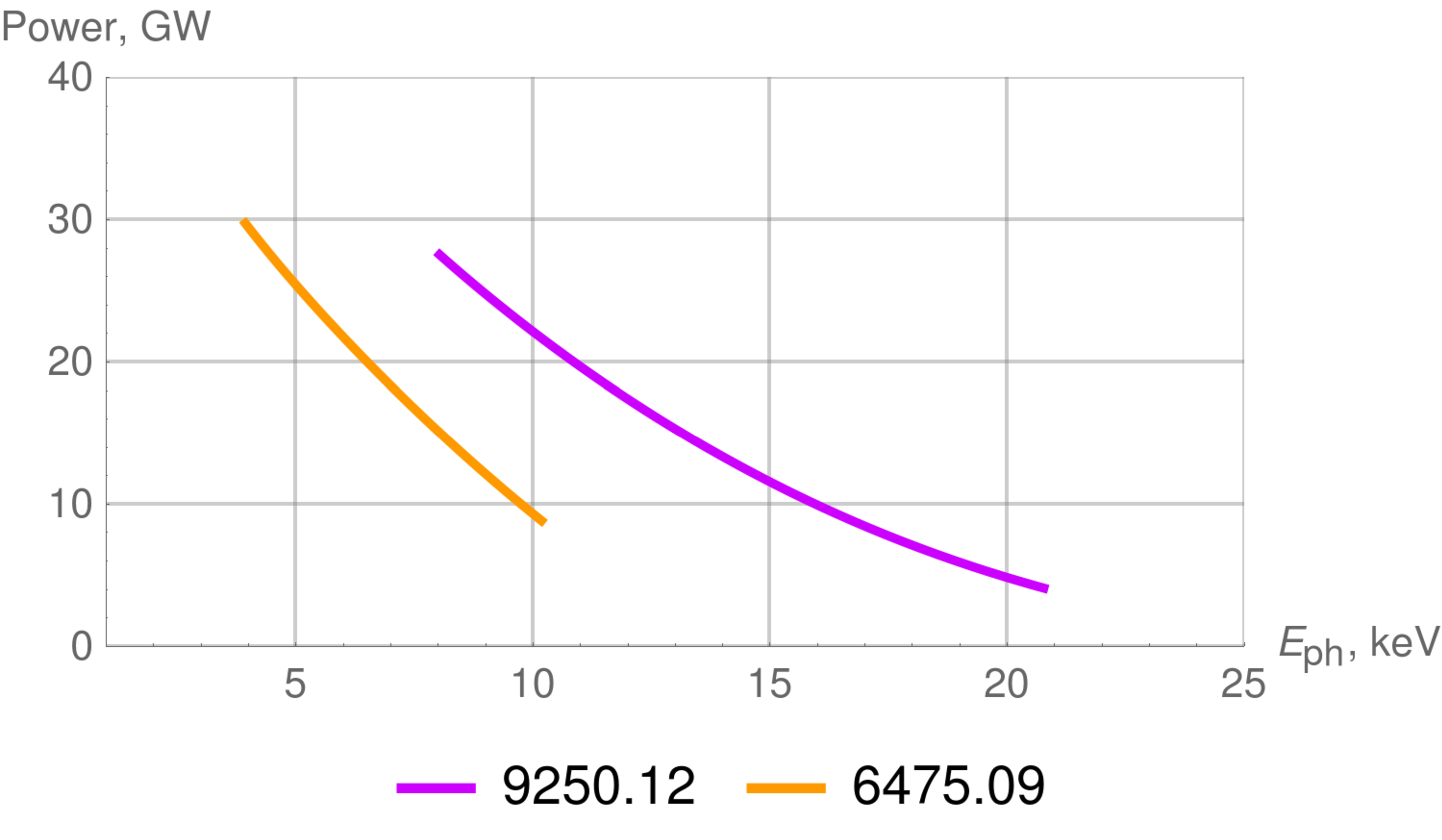}
\end{figure}

We consider first the SASE undulator and evaluate the power gain length and peak power for two different energies and undulator parameter varying between 1 and 2.4. The results, obtained using the Ming Xie code \cite{XieFormulas,Xie:2000kd}, and for the electron beam parameters of Table \ref{beam-parameters} are shown in Figs. \ref{Figure3},\ref{Figure4}.

\begin{table}
\label{lcls-beam-parameters}
\caption{Beam parameters of the LCLS copper linac.}
\begin{tabular}{llcr}      % Alignment for each cell: l=left, c=center, r=right
 Parameter   & Value       \\
\hline
 Electron beam energy, $E$      & 2.5-15 GeV          \\
 Electron bunch charge, $Q_b$ & 0.02 - 0.3 nC\\
 Final rms bunch length, $\sigma_z$& 0.5-52 $\mu$m\\
  Peak Current, $I_{pk}$      & 0.5-4.5 kA      \\
 Normalized transverse emittance, $\gamma \epsilon_\perp$   & 0.2 -0.7 $\mu$m      \\
 Energy spread, $\sigma_E$   & 2 MeV      \\
 Slice energy spread (rms),$\sigma_{E_s}$& 500-2000 keV
\end{tabular}
\end{table}

\begin{table}
\label{lcls-undulator-parameters}
\caption{LCLS-II undulator parameters.}
\begin{tabular}{llcr}      % Alignment for each cell: l=left, c=center, r=right
 Parameter   & SXU Values & HXR Values       \\
\hline
 Undulator period, $\lambda_u$      & 39 mm & 26 mm         \\
 Segment length & 3.4 m&3.4 m\\
 Number of effective periods per segment, $N_p$ & 87 & 130\\
  Minimum operating gap     & 7.2 mm & 7.2 mm     \\
 Maximum $K_{eff}$   & 5.48 & 2.44      \\
  Maximum operating gap     & 22 mm & 20 mm     \\
 Minimum $K_{eff}$   & 1.24 & 0.44      \\
\end{tabular}
\end{table}

\begin{table}
\label{beam-parameters}
\caption{Beam parameters for the double bunch FEL performance evaluation at 4-8 keV.}
\begin{tabular}{llcr}      % Alignment for each cell: l=left, c=center, r=right
 Parameter   & Value       \\
\hline
 Electron beam energy, $E$      & 6.5-9.25 GeV          \\
 Peak Current, $I_{pk}$      & 4 kA      \\
 Normalized transverse emittance, $\gamma \epsilon_\perp$   & 0.4 $\mu$m      \\
 Energy spread, $\sigma_E$   & 2 MeV      \\
Average undulator beta, $\beta_\perp$   & 10 m    \\
Bunch charge, $Q$ & 60 pC \\
Bunch duration, $\tau$ & 15 fs \\
\end{tabular}
\end{table}

\begin{figure}\label{sase-sim}
 \caption{SASE section peak power, bunching, spectrum and energy spread for 4 keV photon energy. The peak output power at the exit of U23 is 6 GW. The red curves are average values over many initial noise distributions.}
 \includegraphics[width=0.41\linewidth]{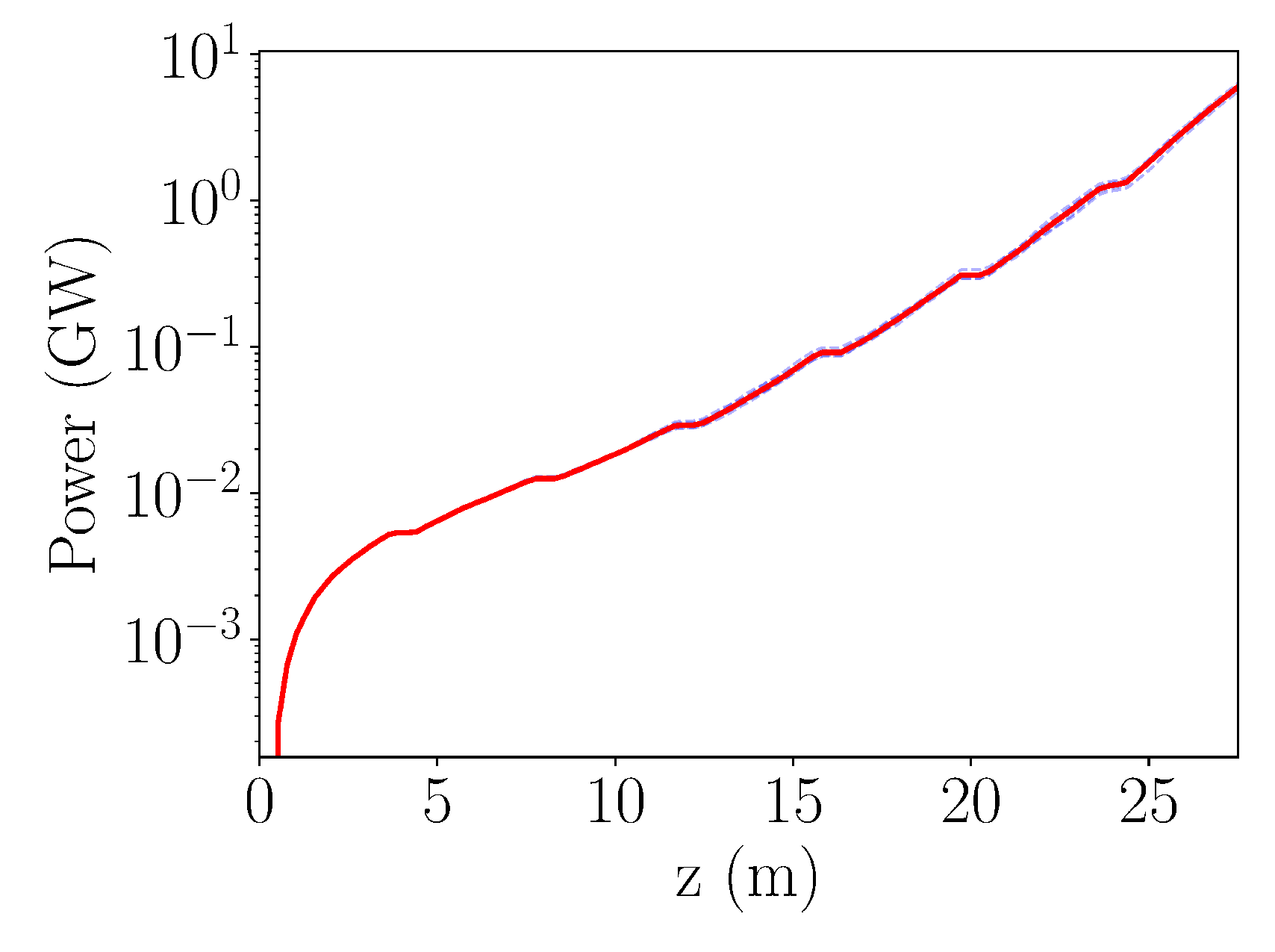}
  \includegraphics[width=0.41\linewidth]{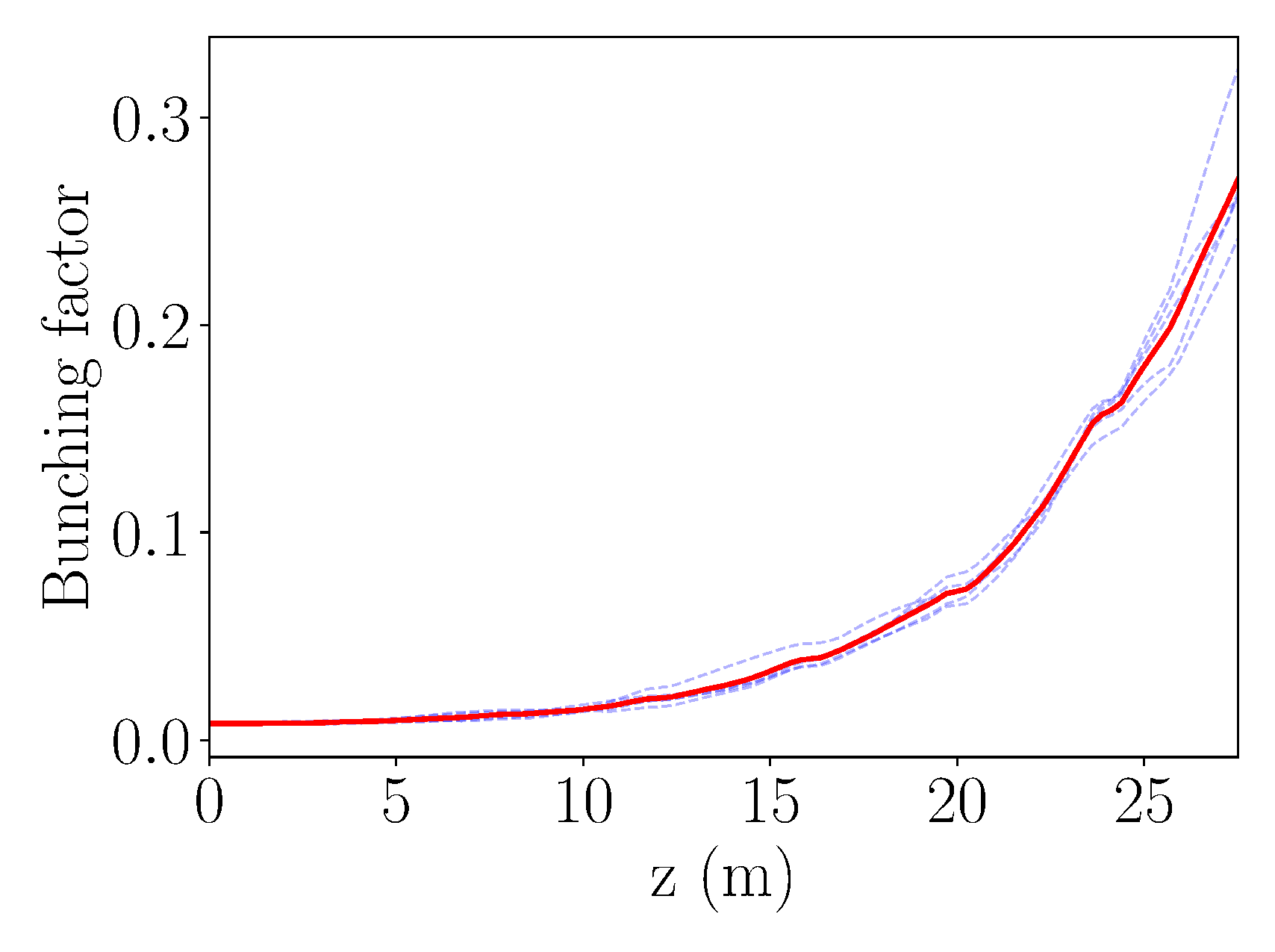}\\
 \,  \includegraphics[width=0.40\linewidth]{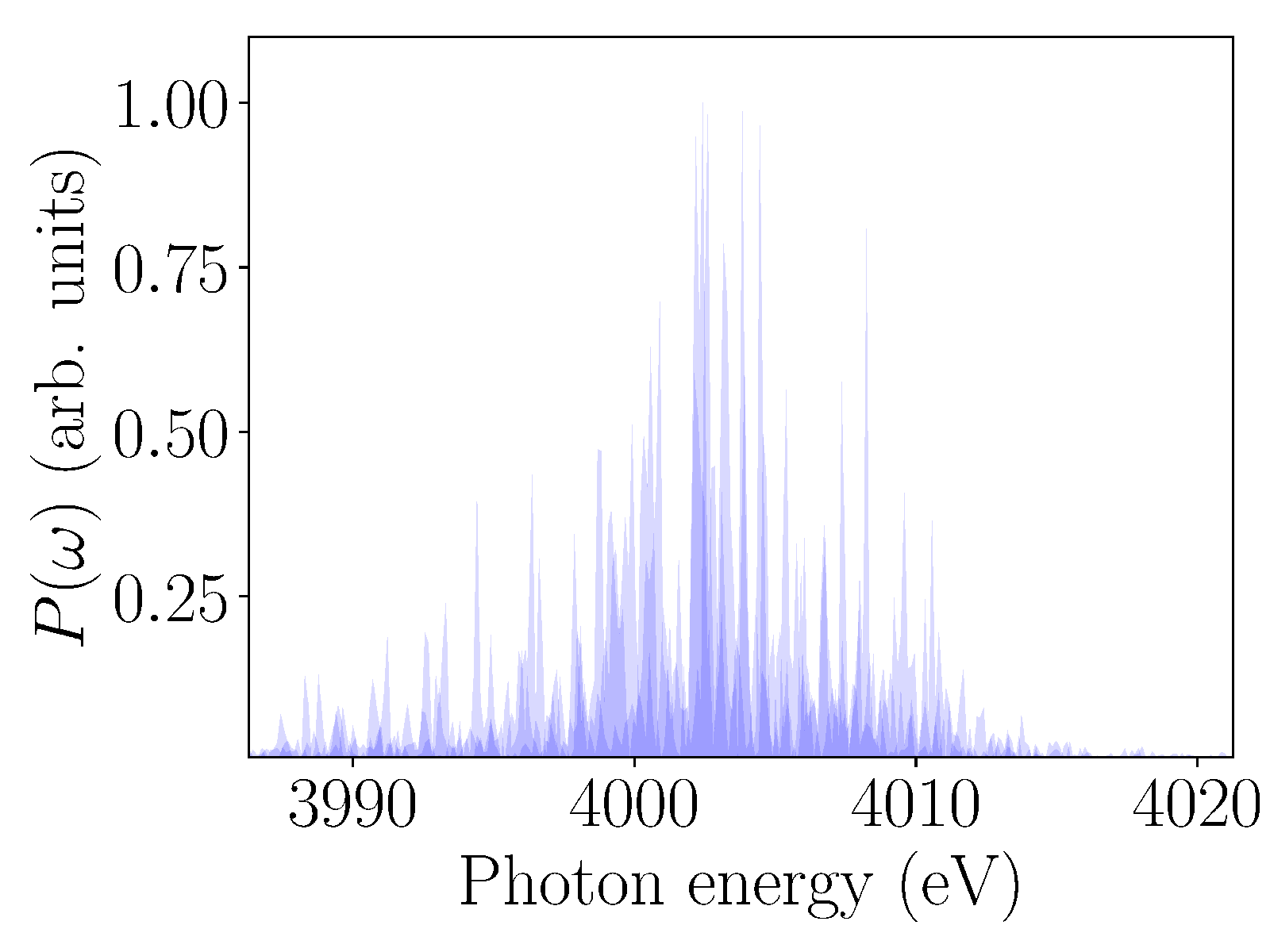} \,
  \includegraphics[width=0.40\linewidth]{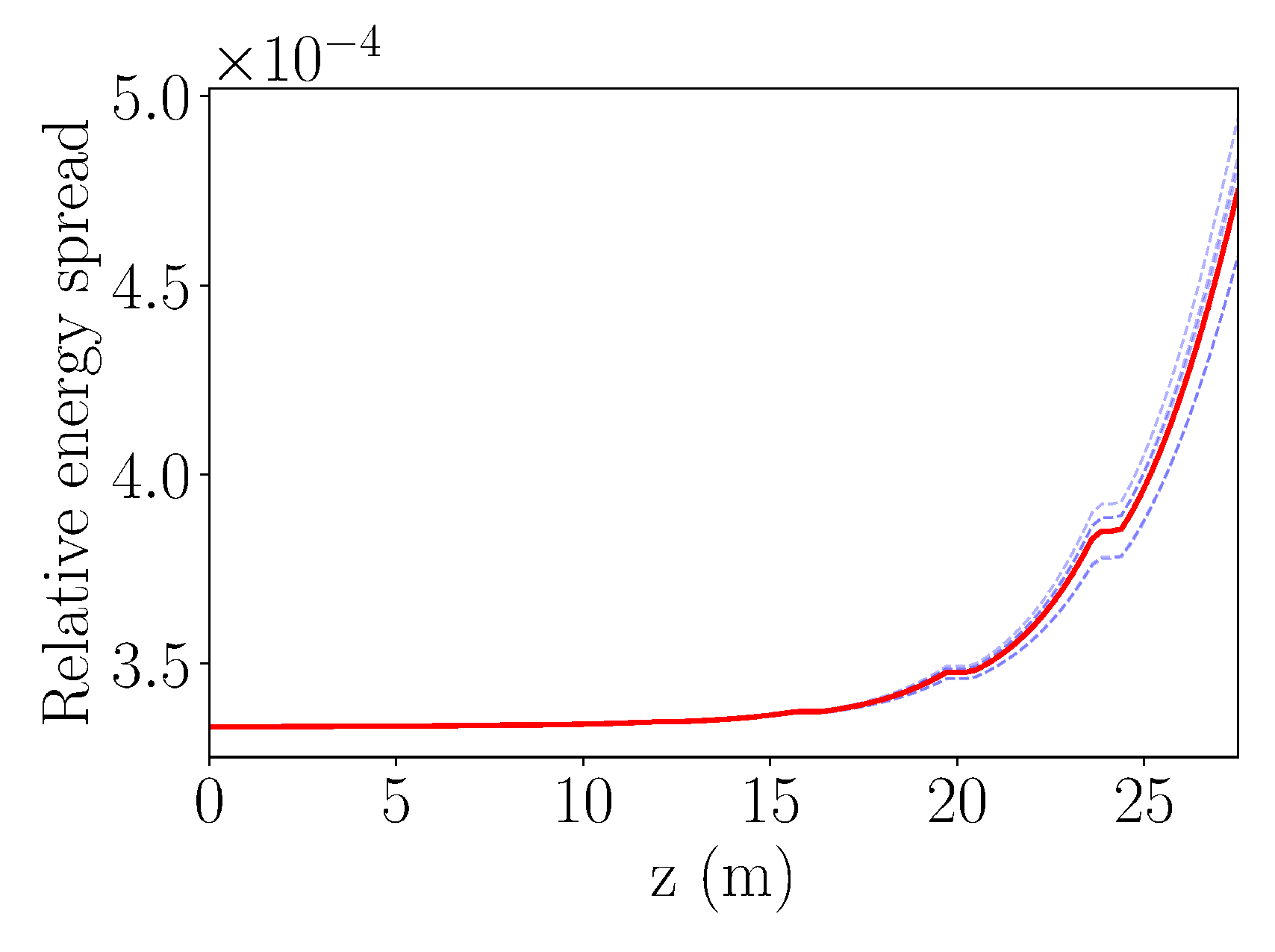}
\end{figure}

\begin{figure}\label{sase-sim-8kev}
 \caption{SASE section peak power, bunching, spectrum and energy spread for 8 keV photon energy. The peak output power at the exit of U23 is 350 MW. The red curves are average values over many initial noise distributions.}
 \includegraphics[width=0.41\linewidth]{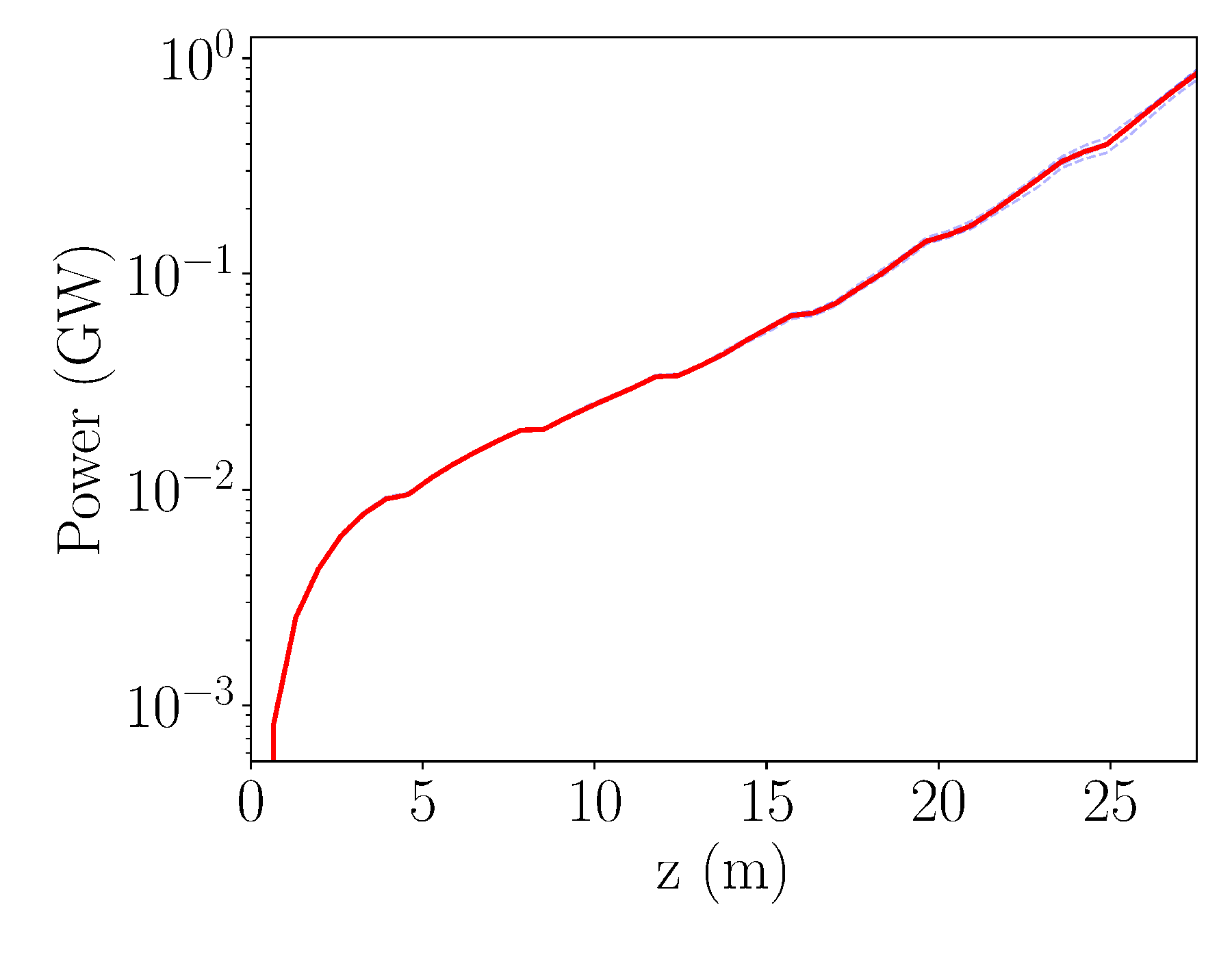}
  \includegraphics[width=0.41\linewidth]{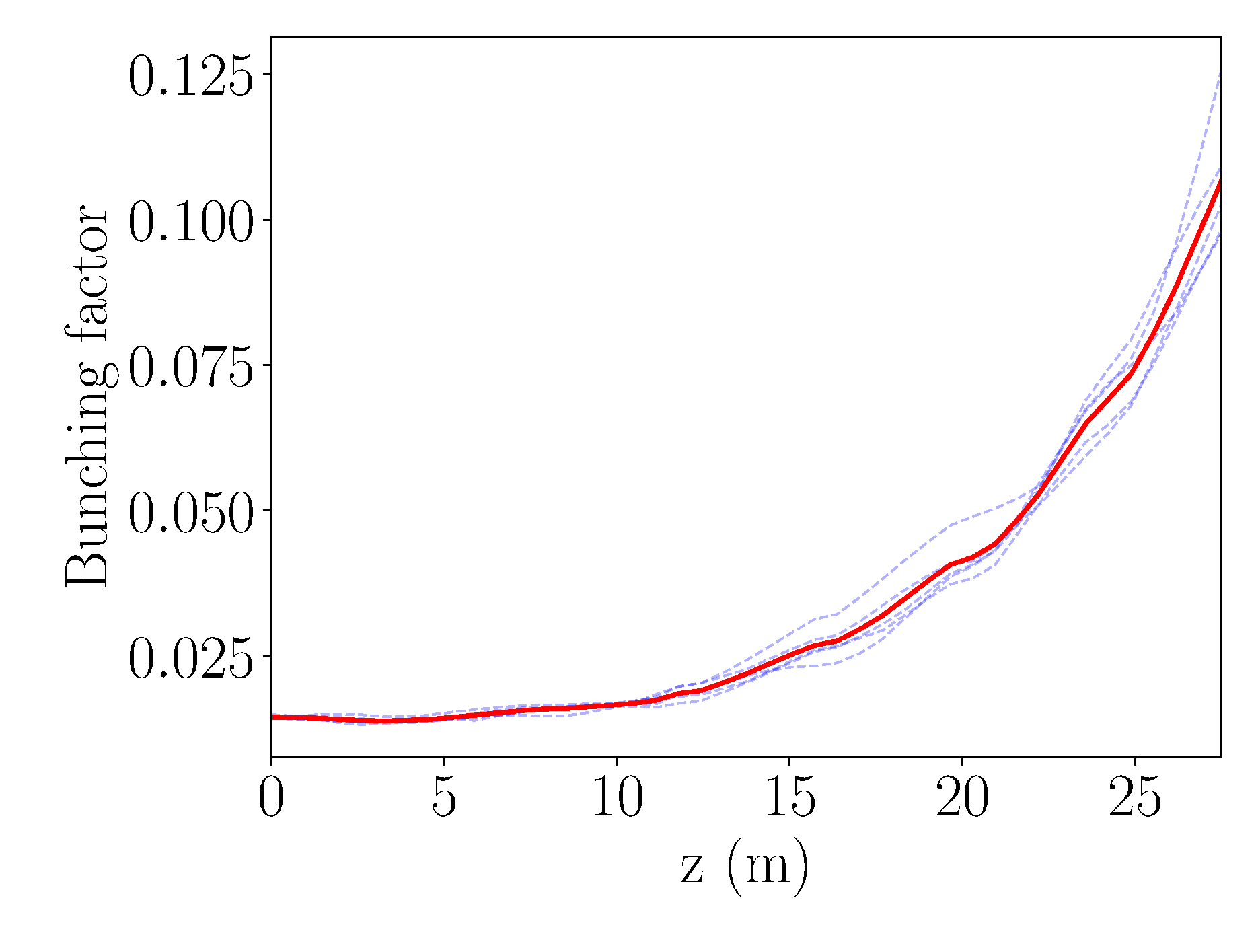}\\
 \,  \includegraphics[width=0.40\linewidth]{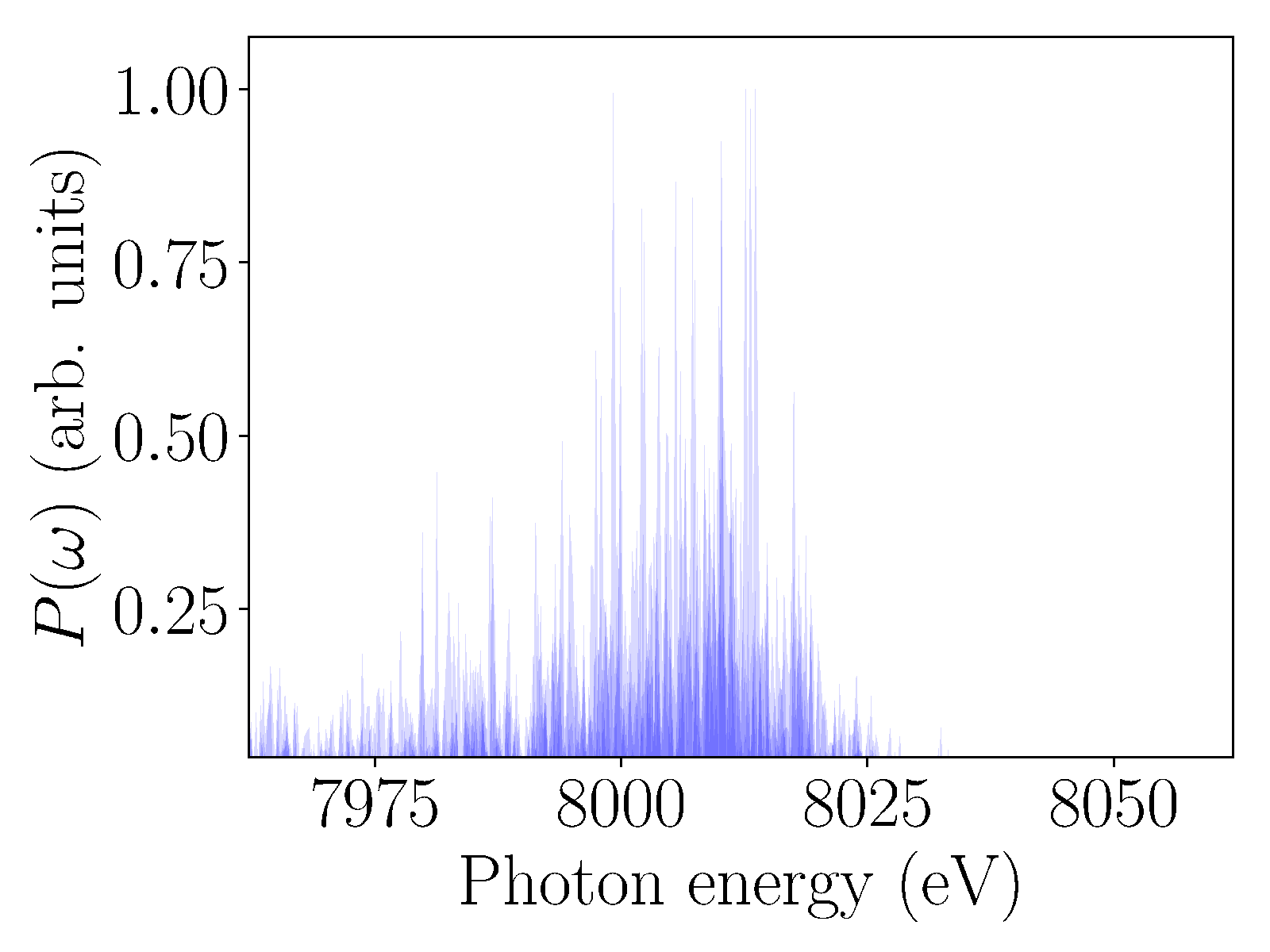} \,
  \includegraphics[width=0.39\linewidth]{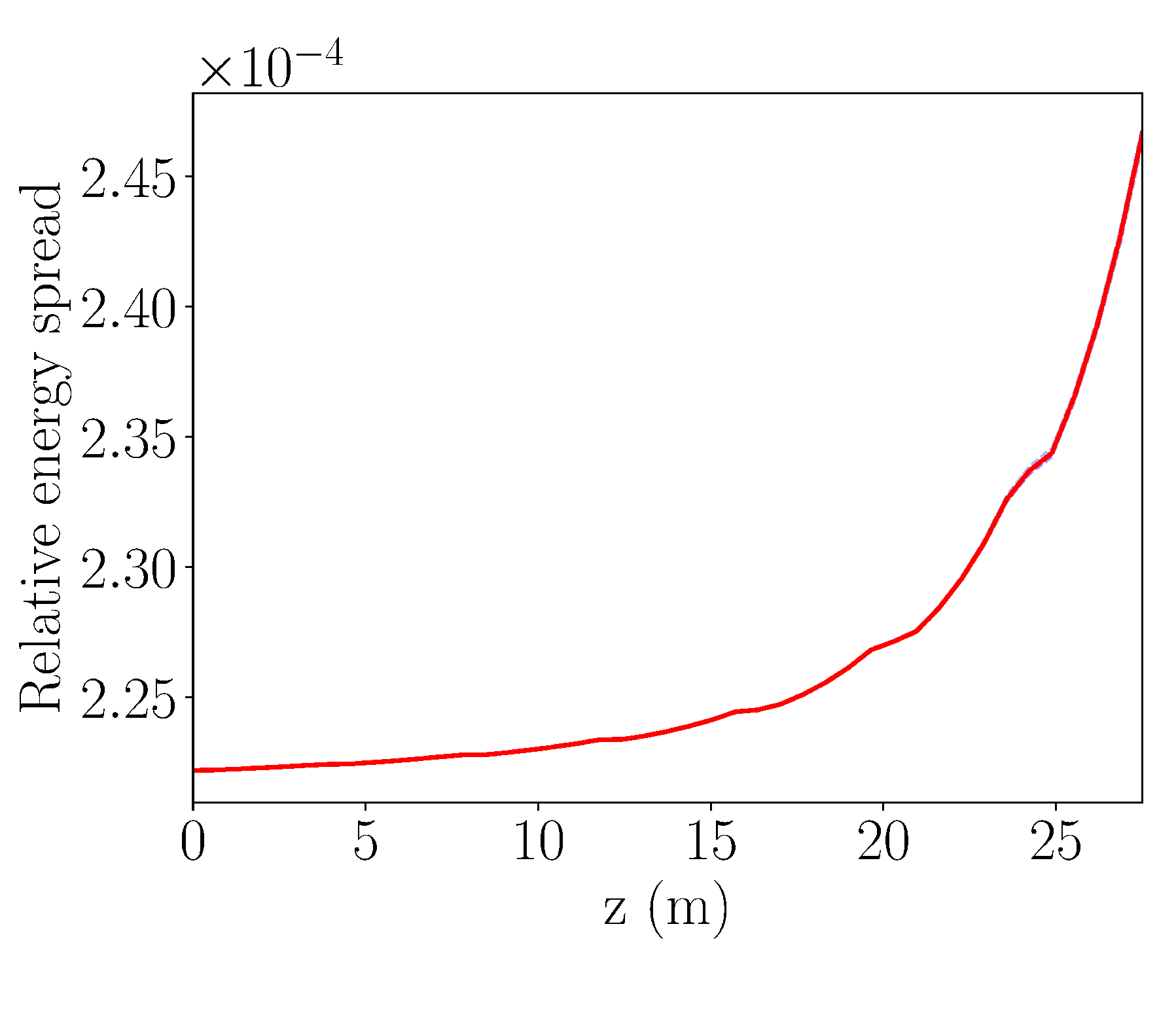}
\end{figure}

Using the first seven undulator sections to generate the seed, the useful undulator length is 23.8 m. 
It is possible to extend the photon energy range where SASE saturation is reached by moving the monochromator to section U27 or later; see Figs. \ref{Figure3}, \ref{Figure4}. However, in this paper, we first discuss LCLS-II performance without making any hardware changes and considering initially lasing in the range 4-8 keV, without reaching saturation in the initial SASE undulator section. The peak power profile, bunching and energy spread, in the first 7 SASE undulator sections, U17 to U23, is obtained by running GENESIS in the time-dependent mode; see Figs. \ref{sase-sim}, \ref{sase-sim-8kev}.
 
The peak power at the SASE undulator exit, which was
used to evaluate the seed signal, is 6 GW at 4 keV 
and 350 MW at 8keV, as shown in Figs. \ref{sase-sim}, \ref{sase-sim-8kev}. 
An alternative setup where the monochromator is located at 
U32 section, as shown in Fig. \ref{lcls2}, which can be used to increase the SASE power at saturation, and therefore provide a much larger seed signal.
In this paper, we mainly discuss the first case, which requires the fewest modifications to the present LCLS-II design.
 
\section{Undulator tapering strategy}
\label{section-taper}
In this section, we discuss how to optimize the tapering of the magnetic field in the amplifier section of the undulator, in order to obtain a large energy transfer from the electron beam to the X-ray pulse. We note that it has been the subject of many studies since the seminal work of KMR \cite{KMR}.

The magnetic field and the resonant phase $\psi_r$ are adjusted in sections U25 to U50 to extract the maximum power using a local step-by-step optimization method.  The resonant phase $\psi_r$, undulator parameter $K$ and beam energy $\gamma m c^2$ are related by
\begin{equation}
 mc^2\frac{d\gamma}{dz} = - \frac{eEK}{\gamma}\sin{\psi_r},
\end{equation}
where $E$ is the electric field acting on the electron.
The beam energy and undulator parameter are also related by the synchronism condition
\begin{equation}
 \lambda = \frac{\lambda_U(1+K^2/2)}{2\gamma^2},
\end{equation}
where $\lambda$ is the photon wavelength and $\lambda_U$ is the undulator period.
 The approach described here focuses on an a-priori selection of the resonant phase profile along the tapered section of the undulator. With a pre-determined variation of the resonant phase, the change in the magnetic field can be calculated at each $z$-location in the undulator using the relationship \cite{RevModPhys.88.015006}:
\begin{equation}\label{tapering}
 \frac{dK}{dz} = -\frac{e}{mc^2} \frac{2\lambda}{\lambda_U}JJ(z)E(z)\sin{\psi_r},
\end{equation}
where $JJ(z)$ is the difference of zeroth and first order Bessel functions
\begin{equation}
 JJ(z) = J_0\left[\frac{a_w^2}{2(1 + a_w^2)} \right] - J_1\left[\frac{a_w^2}{2(1 + a_w^2)} \right]
\end{equation}
and $a_w = K/\sqrt{2}$, is a function of $z$ in the tapered section of the undulator. Here we assume that the average phase and energy of the electrons is the resonant energy and phase.
The algorithm we use consists of computing the approximate numerical solution of Eq. \eqref{tapering}, with the value of the electric field obtained from the GENESIS simulation at each $z$ location. For the n-th integration step, we have
\begin{equation}\label{K-step}
 K_{n+1} = K_n + \alpha_n E_n \sin{\psi_{r,n}},
\end{equation}
where $\alpha_n=-\frac{e}{mc^2} \frac{2\lambda}{\lambda_U}JJ_n(z)\Delta z$.
Since the electrons are distributed across the bunch with nonzero radial extent, the amplitude of the electric field $E$ is approximated as the field amplitude on-axis. Note that this approach is similar to the approach adopted in GINGER's code self-design taper algorithm, which calculates the taper profile at each integration step for a pre-defined constant  resonant phase \cite{WMFawley}. Our method instead allows arbitrary variation of the resonant phase along the undulator. This is similar to the approach discussed in Refs. \cite{PhysRevAccelBeams.20.119902, PhysRevLett.117.174801, JDuris}, but is not limited to express the
resonant phase in the form of a polynomial function, as they assume in their papers.
\begin{figure}
\label{taper_profile}
\caption{An example of an undulator K parameter profile along the undulator length (left panel) and corresponding evolution of the bunching factor (right panel) for 4 keV photons. The undulator parameter changes in range between 1.7 and 1.57. The resonant phase $\psi_r$ profile is plotted with solid line and changes between 0 and 60 degrees.}\includegraphics[width=0.45\linewidth]{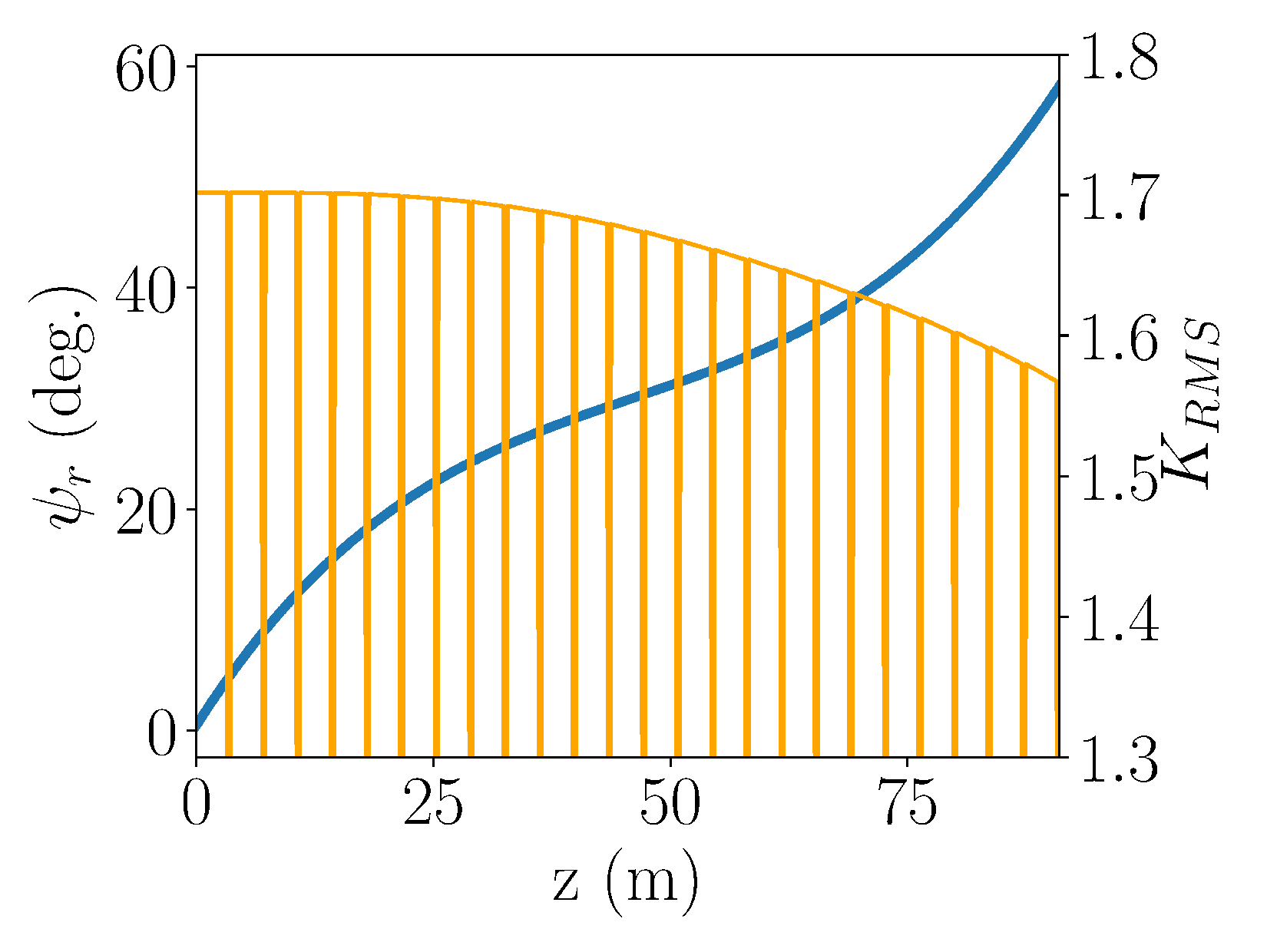}
\includegraphics[width=0.455\linewidth]{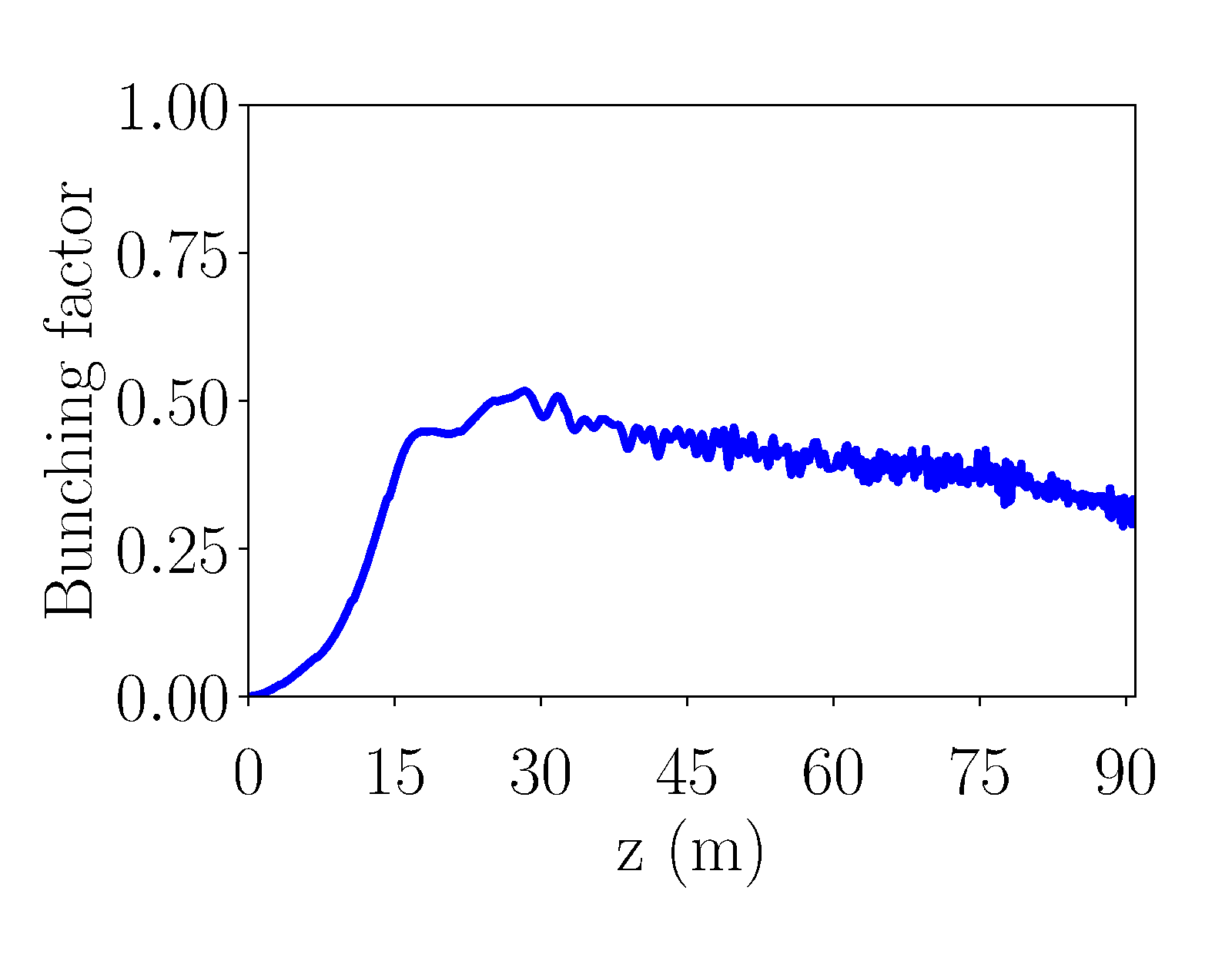}
\end{figure} 
The motivation for allowing arbitrary variation of $\psi_r$ along the undulator is due to the fact that output power depends on the trade-off between the energy loss due to the FEL interaction ($d\gamma/dz \propto \sin{\psi_r}$) and the fraction of electrons trapped $f_t$. In the simplified 1-D limit this can be expressed as $P(z)\propto(f_t \sin{\psi_r})^2$ . This scaling suggests that in the 1-D approximation, the main trade-off when designing a tapered FEL is between the number of electrons trapped in the stable decelerating bucket and the speed at which the trapped electrons lose energy to the radiation field \cite{PhysRevSTAB.18.030705}. This occurs in general because the trapping fraction decreases as the resonant phase and the deceleration gradient increase. The optimal performance is obtained balancing these two effects. 

For the simple case of a constant resonant phase, the optimal value of the resonant phase can be determined analytically and found to be $\psi_r=40$ degrees for a cold electron beam and 20 degrees for a warm beam \cite{Brau,PhysRevAccelBeams.20.110701}. Furthermore, for undulators much longer than the Rayleigh length, the growth of the radiation spot-size during the post-saturation region decreases the effective bucket area in which electrons are trapped and continue to lose energy to the radiation field. These considerations must be taken into account when choosing a particular profile for the resonant phase. 

In general, the resonant phase is chosen to initially follow an almost linear increase, followed by a slow growth around the location of exponential saturation in an undulator. Towards the end of the undulator the resonant phase can be increased more rapidly to extract as much energy as possible from the electrons. Although the trapping fraction decreases, there is no interest in keeping electrons trapped beyond the end of the undulator. An example of the magnetic field change along the undulator and corresponding bunching factor of the second bunch is shown in Fig. \ref{taper_profile}.

\section{The four crystal monochromator}
\label{section-mono}
\begin{figure}
\label{mono}
\caption{Geometry of the four crystal monochromator.}\includegraphics[width=0.7\linewidth]{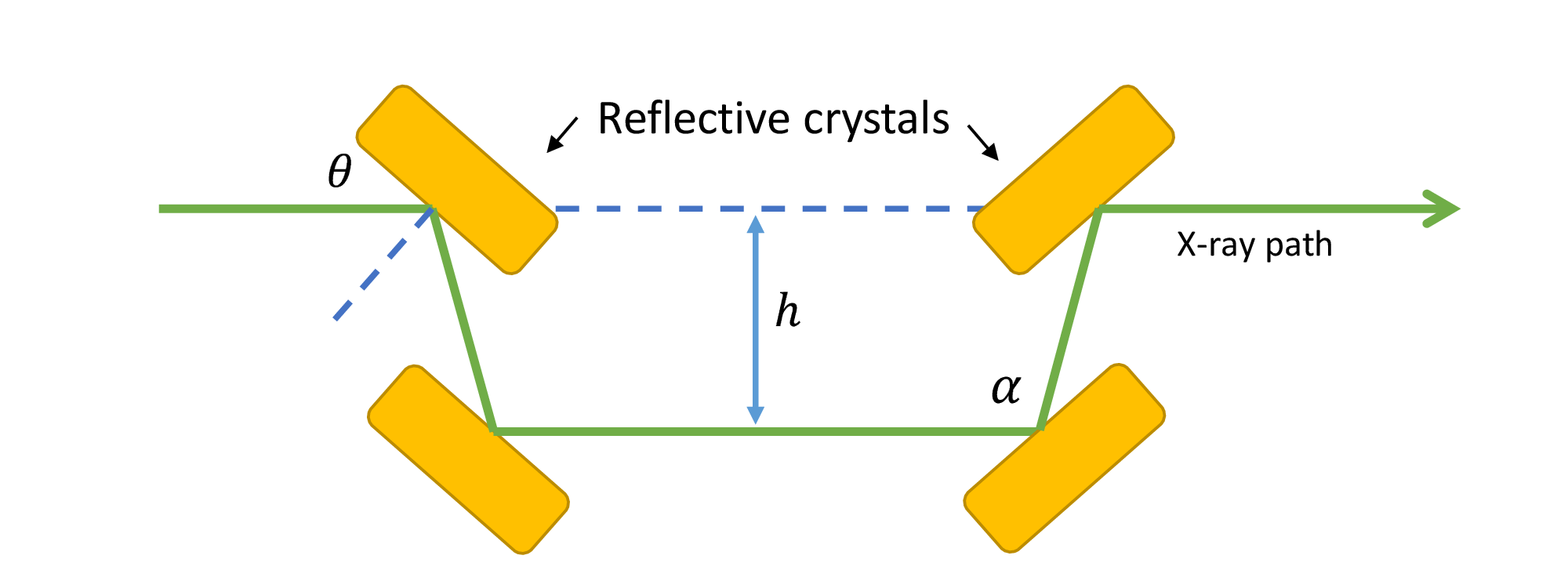}
\end{figure}

\begin{figure}
\label{delay}
\caption{X-ray pulse delay in the four crystal monochromator as a function of the lateral crystal displacement, $h$. For 1 ns we obtain $h=13$ cm at 4 keV (red curve) and $h=37$ cm at 8 keV (green curve).}\includegraphics[width=0.7\linewidth]{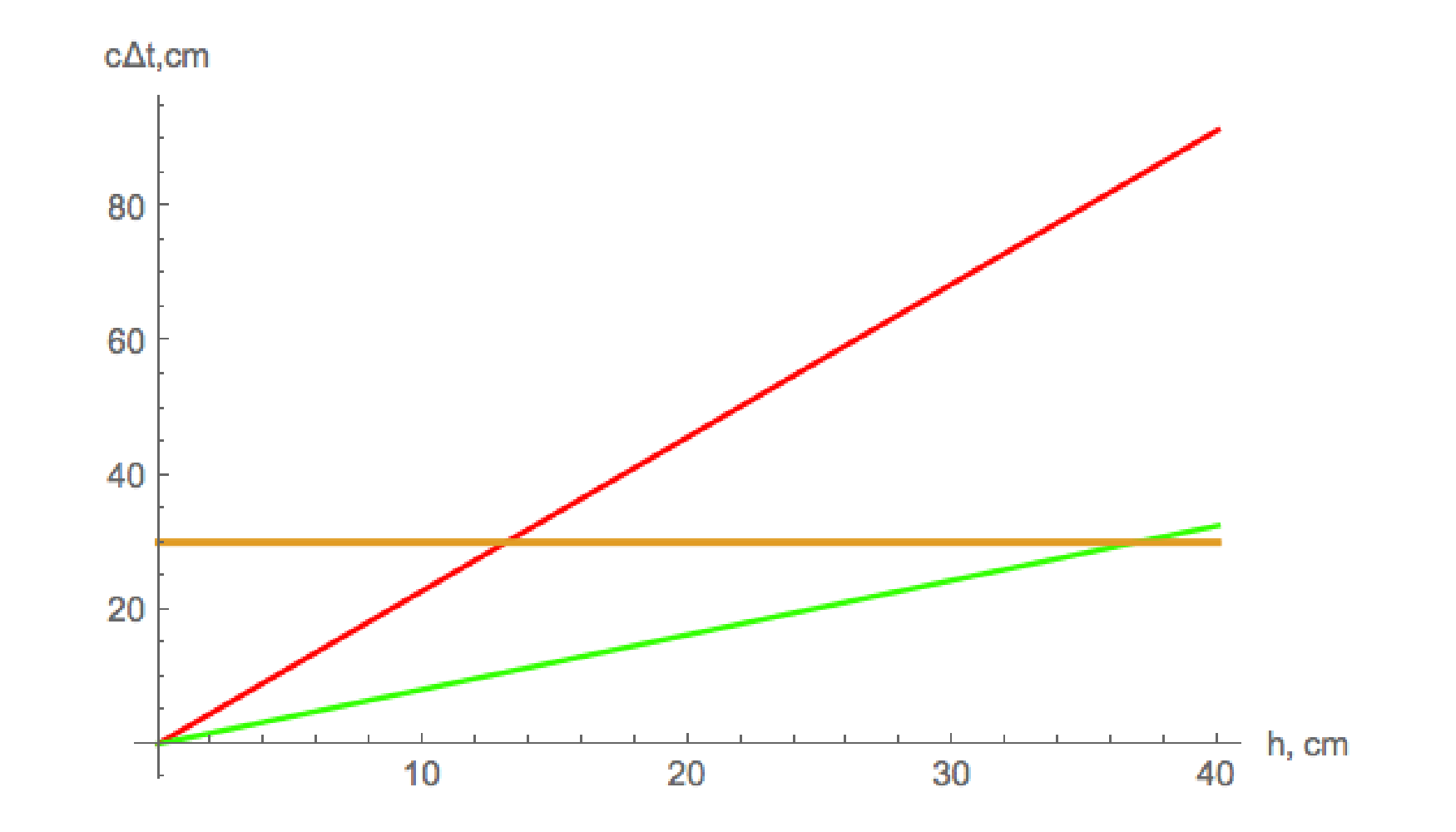}
\end{figure}

The discussion of the four crystal monochromator follows that of reference \cite{EmmaC}, but instead considers photon energies between 4 and 8 keV. The geometry is shown in Fig. \ref{mono}.
The X-ray photons additional path length is given by 
\begin{equation}
 %c\Delta t = 2h(1-\cos{2\theta})/\sin{2\theta},
 c\Delta t = 2h\tan{\theta},
\end{equation}
where $\theta$ is the Bragg angle and $h$ is the lateral displacement.
For our monochromator crystals we choose diamond $(1,1,1)$ crystals. At 4 keV the Bragg angle is $\theta = 48.8$ degrees and the Darwin angle is 14.3 arcsec or about 71.5 $\mu$rad. This gives a bandwidth of $\Delta \lambda / \lambda = \tan^{-1}{\theta}d\theta\approx 7.1\cdot10^{-5}$.  
The evaluation of the X-ray additional path length as a function of the lateral displacement $h$ (see Fig. \ref{mono}) is shown in Fig. \ref{delay}. The reflectivity curves are shown in Figure 10.

An alternative choice of crystal material is Silicon which has twice as large bandwidth as diamond, or Germanium, which is charaterized by very wide reflectivity window comparable to SASE width.
Both choices will result in multiple SASE modes passed to the amplifier and generate broader final spectral content \cite{SASE_sat}. 

The SASE signal bandwidth is about $2\cdot10^{-3}$ (see Fig. \ref{sase-sim}) with 95\% reflectivity and $7\cdot10^{-5}$ bandwidth. The seed power, starting from 6 GW peak power at the exit of the first seven undulator sections, is reduced to 150 MW (efficiency of 2.5\%). The seed power reduction at 8 keV is similar.

\begin{table}\label{crystals}
\caption{Bragg angle, Darwin width and photon energy acceptance for Diamond (1,1,1) at 4 and 8 keV fundamental X-ray photon energy.}
\begin{tabular}{lccccc}      % Alignment for each cell: l=left, c=center, r=right
Diamond (1,1,1) & $E_{ph}$, keV & Bragg Angle & Darwin width, $\mu$rad & $\Delta\omega / \omega$ & Efficiency, \%   \\
\hline
& 4      & 48.8 & 71 & $7.1\cdot10^{-5}$  & 2.5    \\
& 8      & 22.1 & 25 & $6.2\cdot10^{-5}$  & 2.5    \\
\end{tabular}
\end{table}

\begin{figure}
\label{8kev-reflection}
\caption{Diamond (1,1,1) reflectivity curves at 4 keV (left panel) and 8 keV (right panel) photon energy obtained from XOP code \cite{XOPcode}.}
\includegraphics[width=0.45\linewidth]{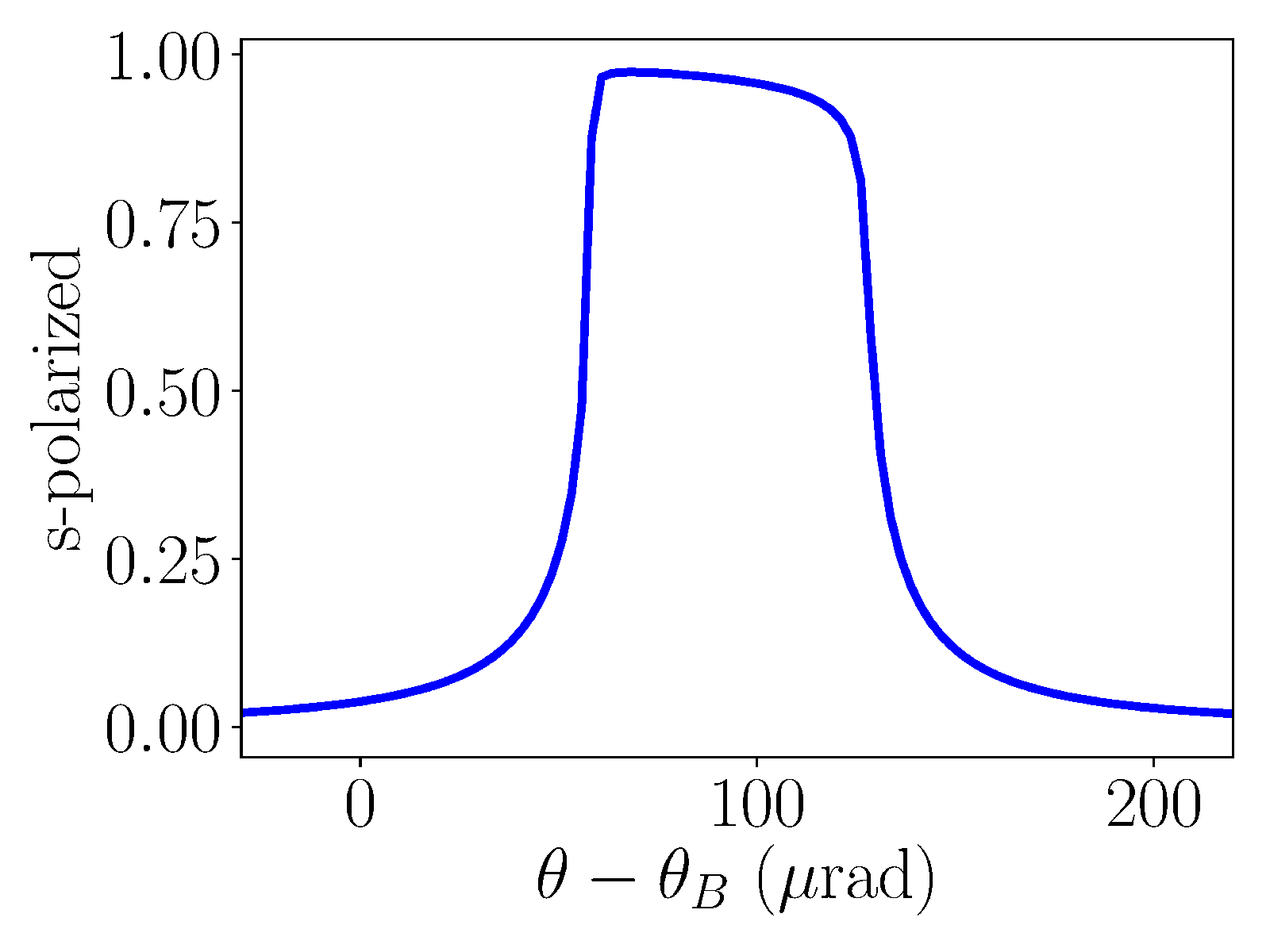}
\includegraphics[width=0.45\linewidth]{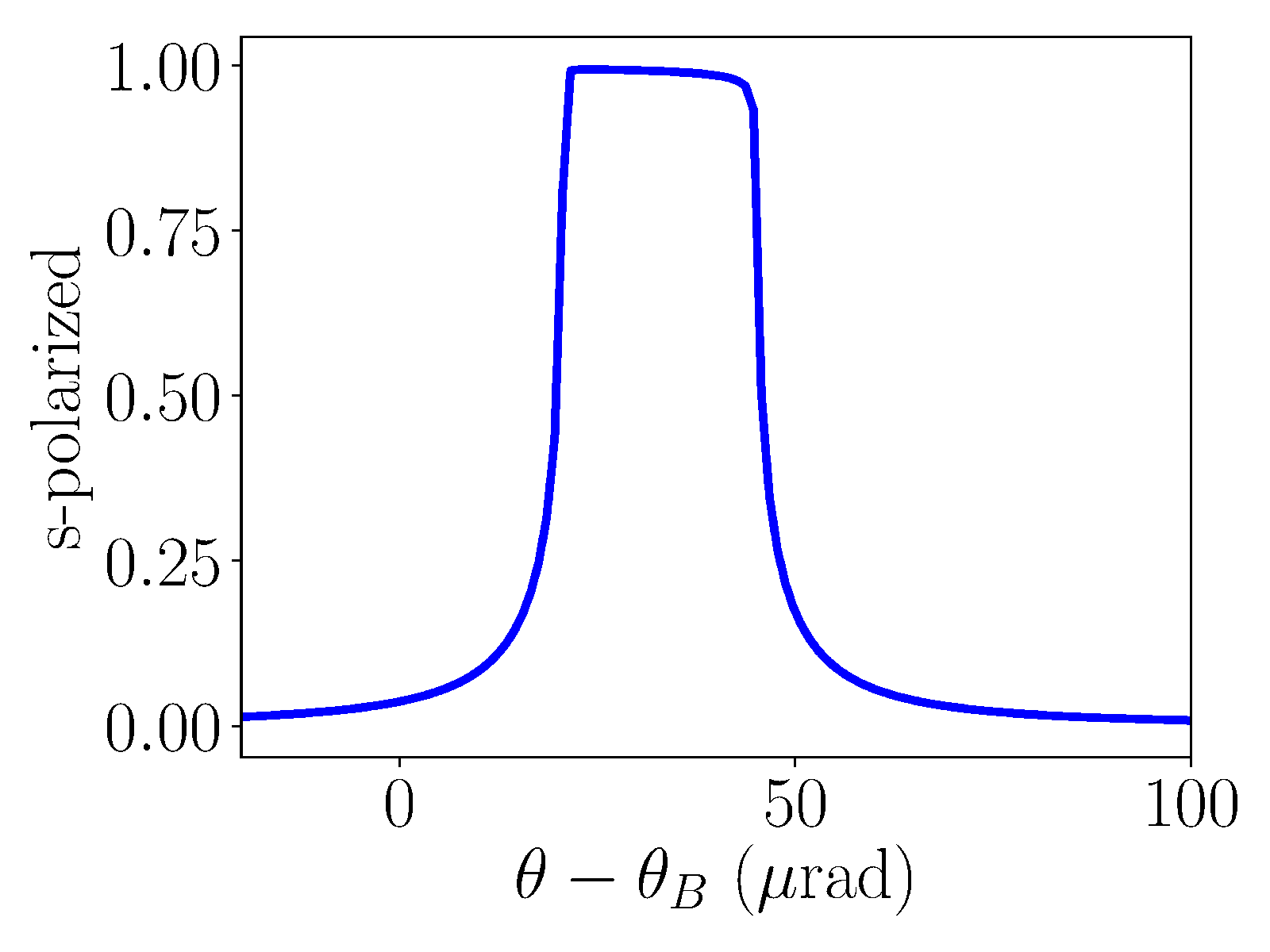}
\end{figure} 

As shown in Tab. \ref{crystals} and Fig. \ref{8kev-reflection}, the four crystal monochromator based on diamond (1,1,1) crystals covers the full energy range from 4 to 8 keV with about the same photon energy acceptance. It provides continued tunability of the X-ray pulse in this energy range by rotating the crystals and simultaneously changing the lateral displacement $h$.
Lower photon energies would require a different choice of crystals.

\section{Present double-bunch LCLS linac operation}
\label{section-double}
The SLAC copper linac driving LCLS normally operates with a single electron bunch per macropulse. 
It has been shown by Decker et al. \cite{FJDecker} that multiple bunches 
can be generated within the linac macro-pulse.
The bunches are separated in time by a multiple of the linac RF
frequency, with small variations useful to control their relative energy. In our study, we consider
two bunches separated by three RF cycles, or 1.05 ns. The bunches are created by sending two
light pulses from two independent lasers on the LCLS photoinjector cathode.
Their relative charge difference
can be controlled to about 1\% level and their individual time separation can be adjusted with a precision of 0.07 ps.
Longitudinal and transverse wakefields generated by the first bunch act on the successive bunch.
The beam loading (or longitudinal wakefield) is 70 V/pC/m. For 1 km long RF linac and 60 pC bunch charge,
we expect the second bunch to be 4 MeV lower in energy, or 0.07\% at 6 GeV beam energy.
This can
be compensated by having a 0.08$^\circ$ phase difference between the two bunches in second section, L2, of the linac (6 GeV *
$(\cos {35^\circ}-\cos{35.08^\circ})$ = 4 MeV). The difference of 0.08\% is also compensated by timing the global RF
pulse, since 0.08\% is about the ratio of the 1.05 ns separation divided by the 825 ns RF fill time.
The transverse wake field could be used to give the second bunch a kick to oscillate around the axis, as needed in the DBFEL scheme. However for now we assume for simplicity to use a separate transverse RF cavity to give the transverse kick to the second bunch and to compensate the linac wakefield if needed.
The transverse effects are strong and can reach orbit differences of 100 $\mu$m in the undulator,
which would inhibit lasing of the second bunch if not corrected, see Fig. 3 in \cite{Decker:2018oot}.
This separation due to wakefields can be used to adjust it to the desired transverse separation.
Successful experiments have been done using two bunches, such as the “probe-probe” method, see Tab. 1 in \cite{Decker:2015gry},
where the photon energy is exactly the same going through a monochromator.

\section{DBFEL performance characteristics}
\label{simulation-results}
In this section we discuss
the characteristics of the X-ray pulse at the seeded amplifier exit,
for different photon energies, as a function of the seed power. Our study is based on
numerical simulations using the 3D time-dependent code GENESIS. Here
we considered the LCLS-II HXR undulator with a step of 5 undulator periods, evaluating $K$ using Eq. \eqref{K-step}.
The X-ray seeded amplifier power output and spectrum are evaluated for the cases of an initial seed signal
equivalent to the SASE noise, 10 kW, the case of a single electron bunch, and using a DBFEL.
The SASE noise signal of about 10 kW represents the case when no monochromator is inserted, and hence provides a baseline of X-ray power for selected tapering scheme.

We evaluated DBFEL performance for 4 keV and 8 keV photon production, the two extremes of our range of interest. For our studies, we selected the resonance phase profile shown in Fig. \ref{taper_profile}, which can be analytically approximated by $\psi_r(z) = 1.3538 z - 0.0231 z^2 +  0.00017 z^3$.
Other ways to optimize the resonant phase profile have been discussed in \cite{WU201756,Wu:2018tdh,TSAI2018}. Deep multi-objective optimization of the DBFEL scheme will be the main focus of a separate study. Hereafter we discuss the power output of the DBFEL based on our tapering strategy.

Additionally, we also compare the power output with that of an FEL operating in the 1D
regime, driven by an electron beam with negligible energy spread, the most favorable case, given by \cite{Juhao}:
\begin{equation}\label{power-coherent}
 P_{coh}(z) = \frac{Z_0K^2JJ(z)^2I_{pk}^2b^2z^2}{32\sqrt{2}\pi\sigma^2\gamma^2},
\end{equation}
where $I_{pk}$ is the peak current and $b$ is the bunching factor.
In this equation one assumes peak current dependent on $z$ as $I_{pk} = I_0 f_t (z)$, where $f_t(z)$ is the trapping fraction. 

\subsection{4 keV photon case}

Using the results of sections \ref{section-dbfel} and \ref{section-mono} the seed power can
be as high as 150 MW when using two bunches and the four
crystals monochromator. When we consider a single bunch
the seed power is limited to 5 MW to avoid an additional energy spread increase. 
\begin{figure}
\label{power-4kev}
\caption{Amplifier undulator peak power output at 4 keV as a function of $z$ compared to different seed power signals: 10 kW corresponds to the SASE noise case, 5 MW is the single bunch seed power, and 150MW is the maximum seed power for the two bunches case. Dashed line corresponds to the coherent power value given by Eq. \eqref{power-coherent} for the beam parameters provided in Tab. \ref{beam-parameters}.}\includegraphics[width=0.6\linewidth]{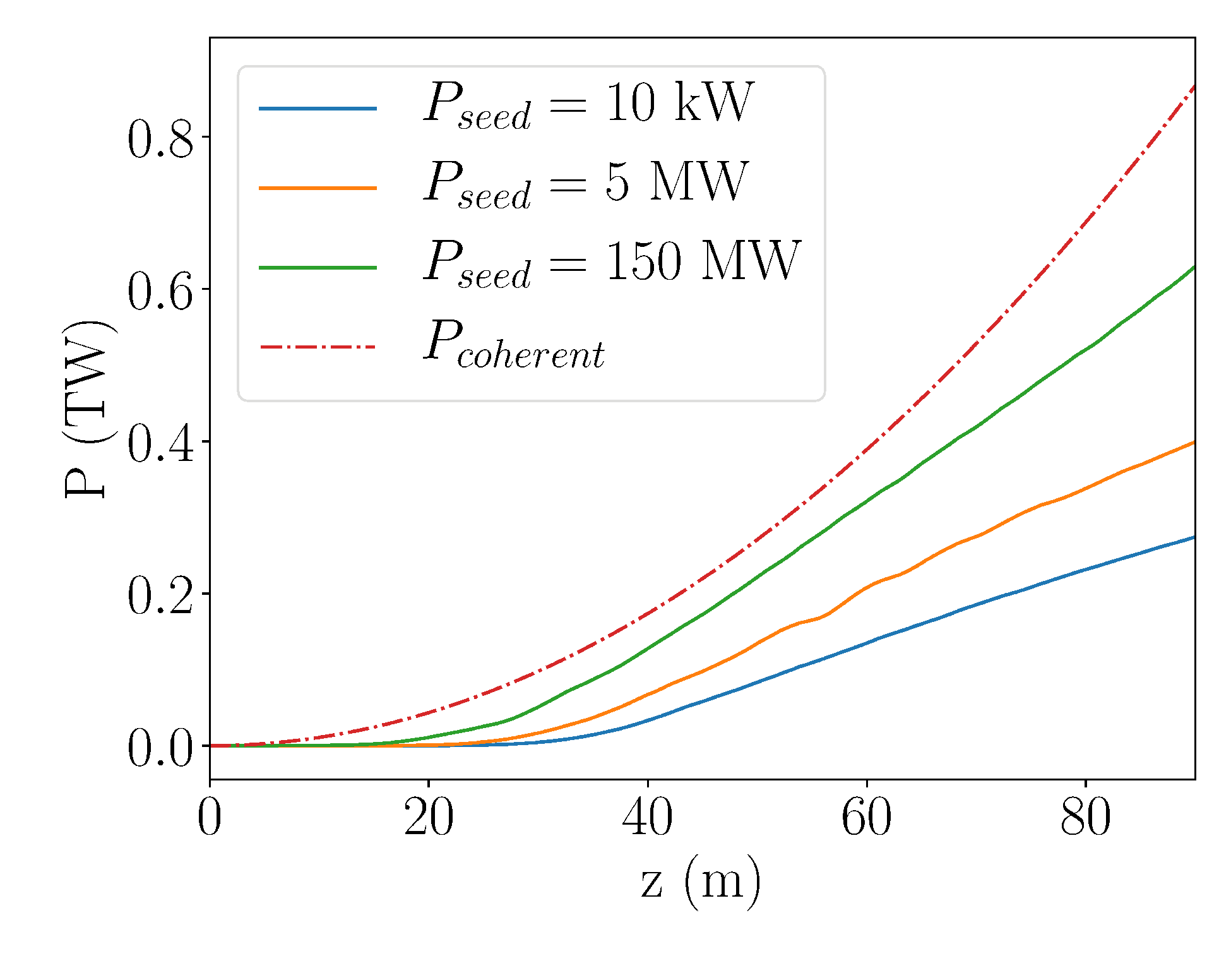}
\end{figure} 
The performance of the DBFEL for 4 keV photon production is presented in Fig. \ref{power-4kev}. For the case of DBFEL we obtain 650 GW peak power downstream of the amplifier, which for the flat-top bunch with a duration of 15 fs yields about 10 mJ peak energy. For signle bunch case the power is two times smaller, 320 GW. 
The power spectra for the two cases and the power temporal profile along the bunch are given in Fig. \ref{4kev-spectra}. 
Notice, that we have a flat profile along the bunch, following our assumption, discussed in Section \ref{section-dbfel}, that the bunch current profile that we generate is flat. The noise present in the power distribution along the bunch is due to the growth of the SASE signal along the undulator due to the intrinsic beam noise.

One can see the four crystal monochromator yields a cleaner spectrum with relatively the same bandwidth. The amount of power stored in the fundamental harmonic for the DBFEL case is 92\%, while for the single bunch case it's about 82\%. 
Correspondingly the
temporal profile also improves for the DBFEL case.
In summary, the DBFEL provides X-ray pulses with higher output peak power and more
power stored in the main harmonic.

\begin{figure}\label{4kev-spectra}
 \caption{Power spectrum of 4 keV photons for the case of single bunch (left panel) and the DBFEL (right panel). X-ray power profile in the time domain (bottom panel).}
 \includegraphics[width=0.45\linewidth]{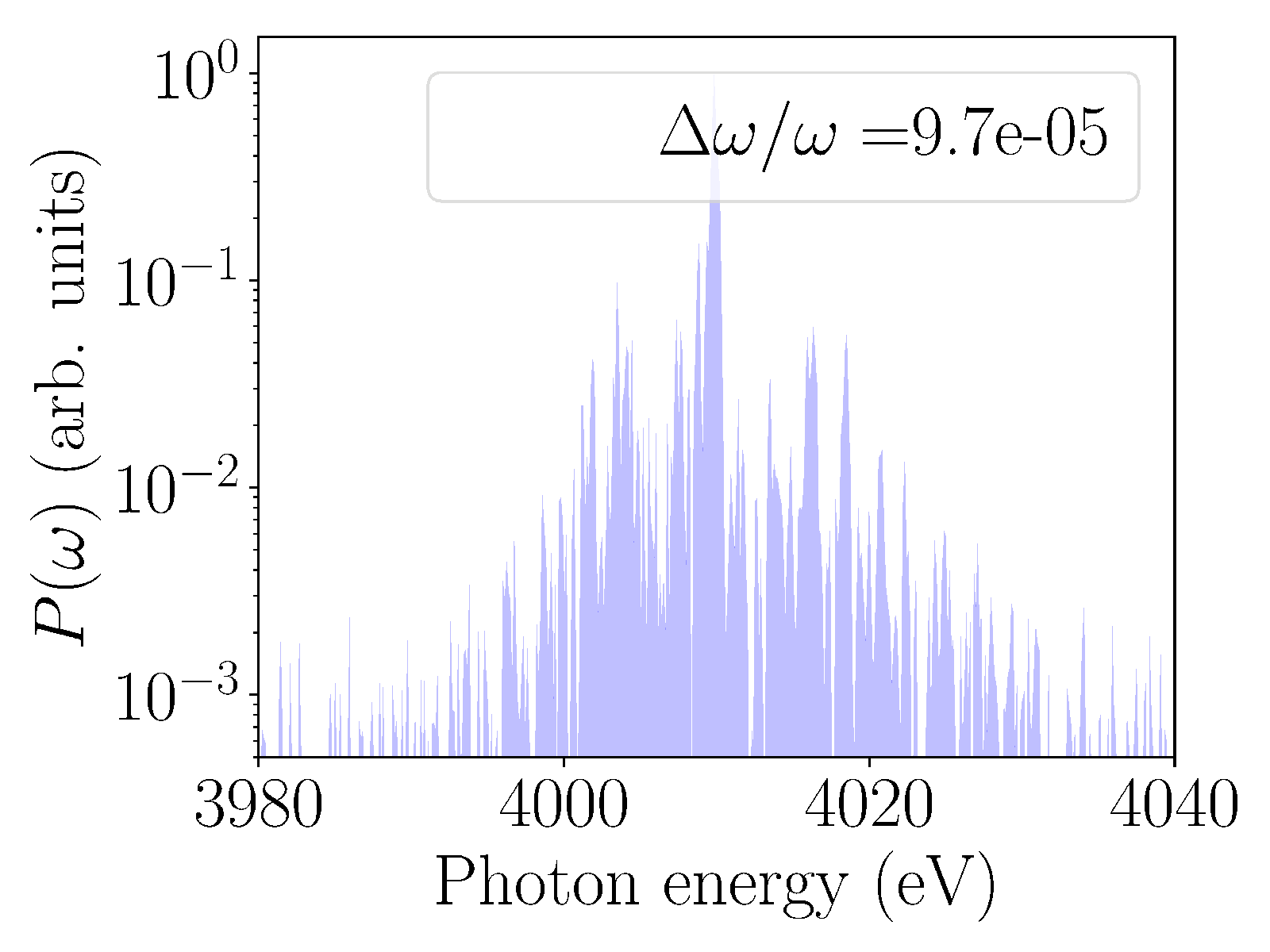}
  \includegraphics[width=0.45\linewidth]{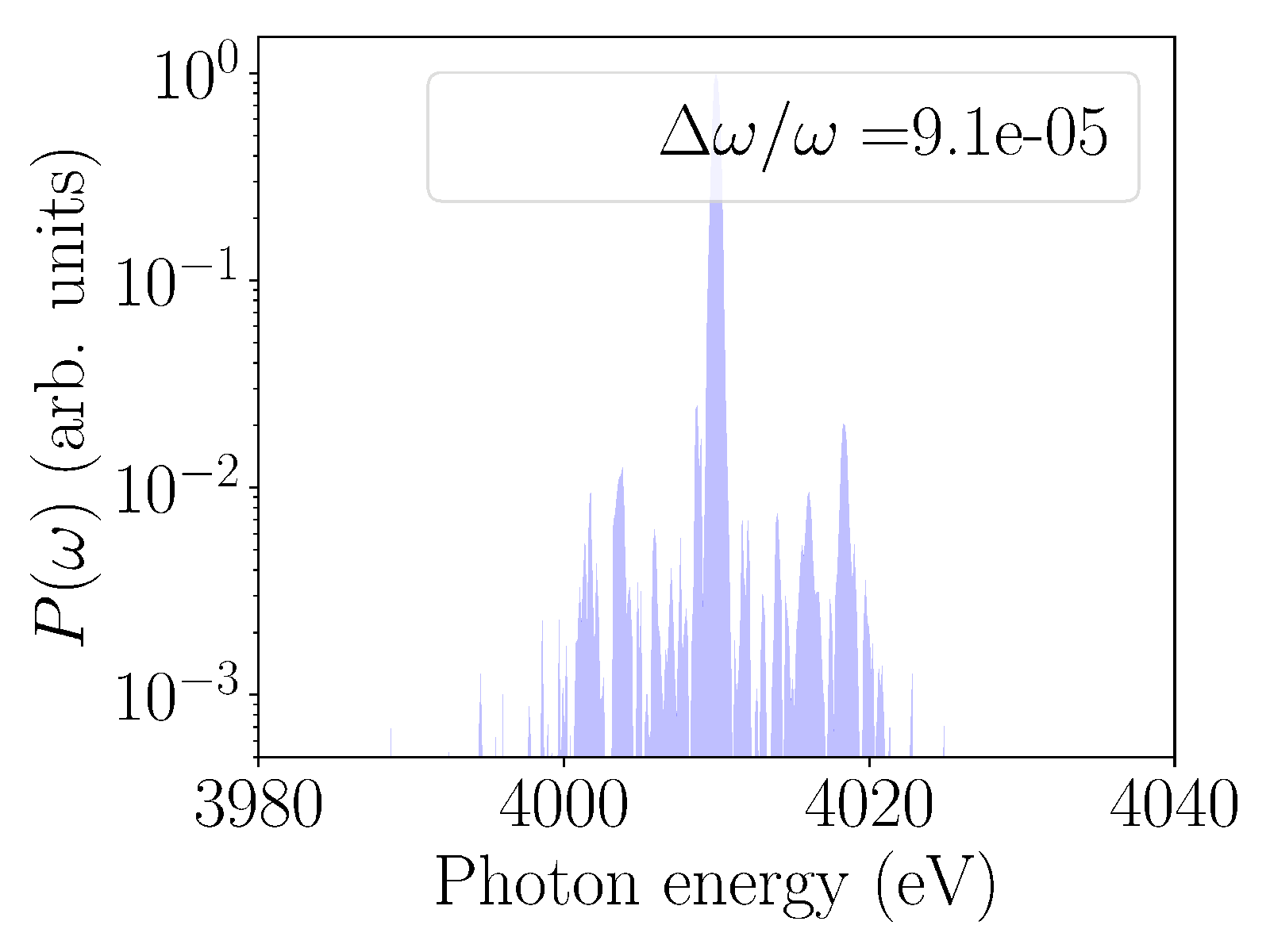}\\
  \includegraphics[width=0.50\linewidth]{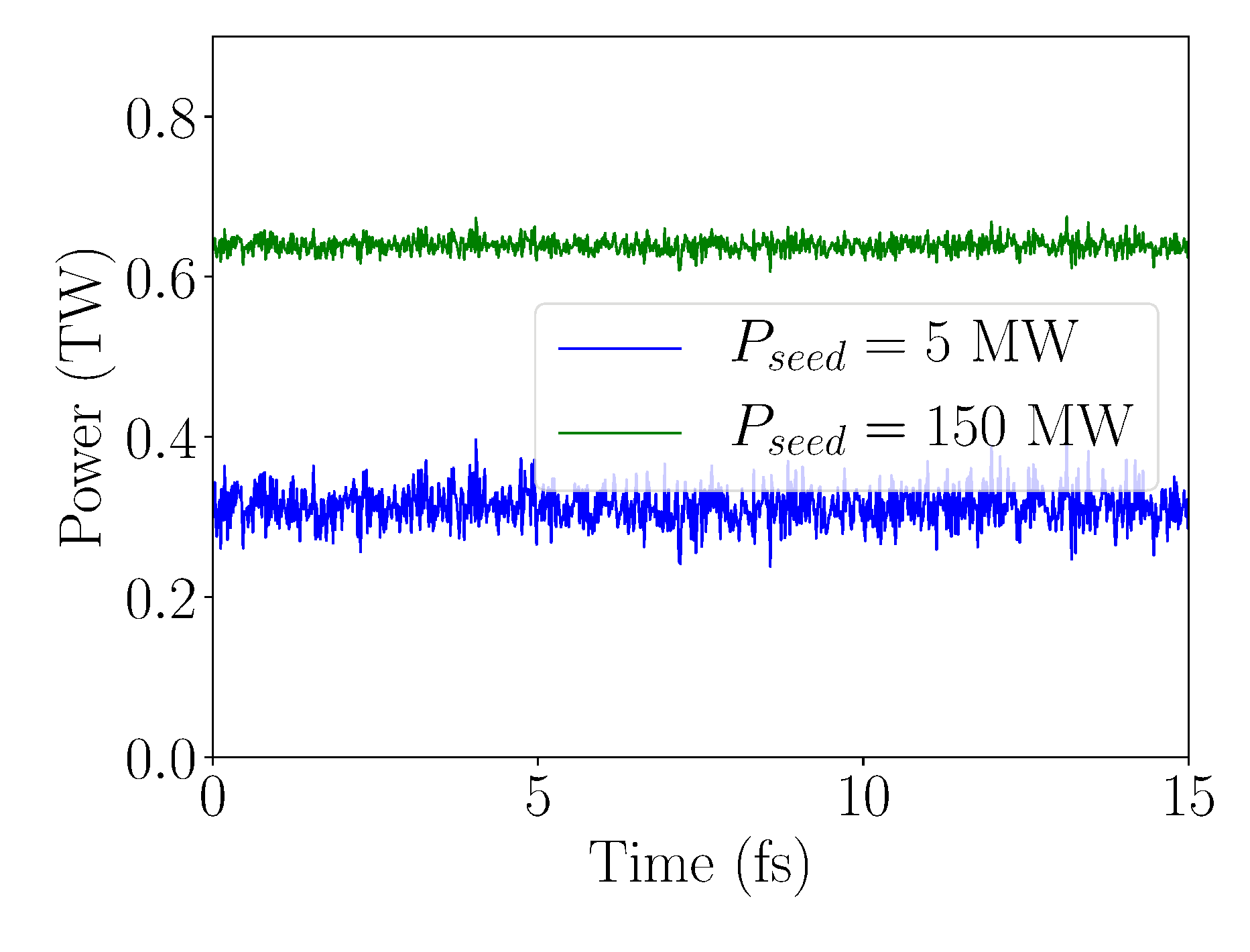}
\end{figure}
 
\begin{figure}\label{4kev-harmonics}
 \caption{Third and fifth harmonic of 4 keV photons power as a function of $z$ in the amplifier undulator for 5MW (left panel) and 150 MW (right panel) input seed power.}
 \includegraphics[width=0.445\linewidth]{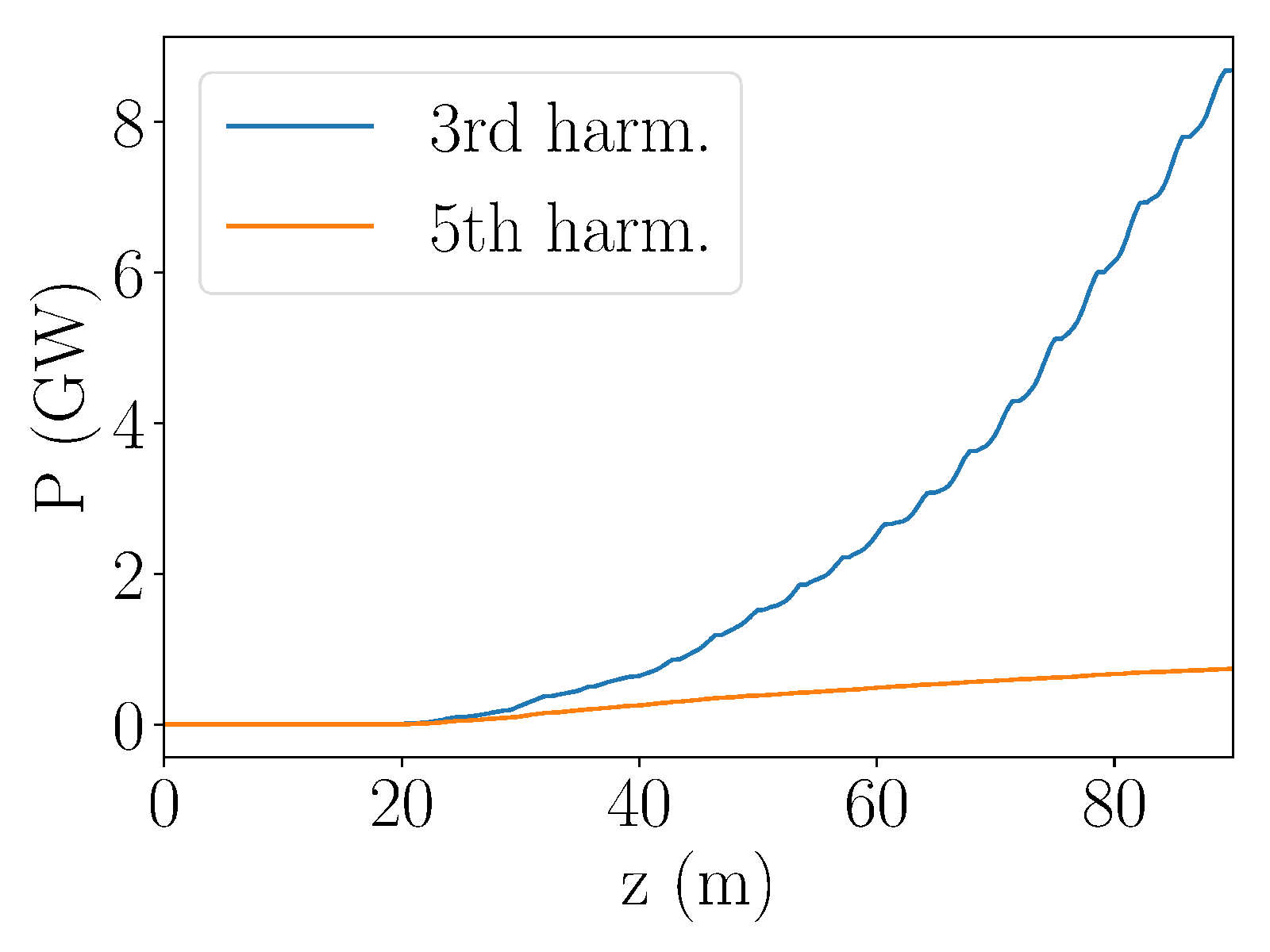}
  \includegraphics[width=0.45\linewidth]{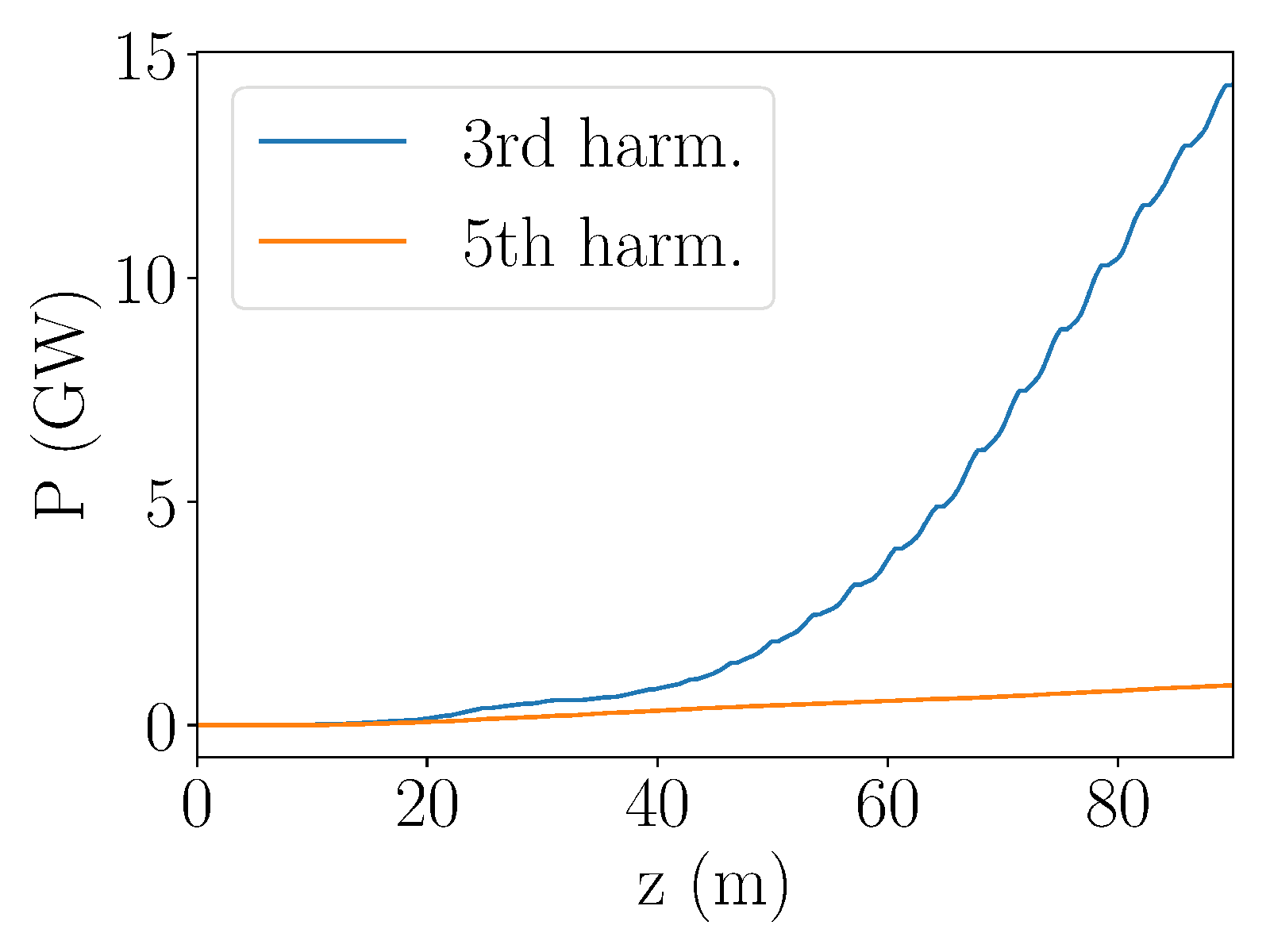}
\end{figure}

The 3-rd and 5-th harmonic of the spectrum obtained from nonlinear harmonic generation are displayed in Fig. \ref{4kev-harmonics}, showing again the advantage of the DBFEL.

\subsection{8 keV photon case}
To establish the upper operating range of the DBFEL we consider the case of 8 keV photon production. 
Placing the four crystals at U24 location the power output at the
SASE section is 350 MW, as shown in Fig. \ref{sase-sim-8kev},  and the seed signal at the amplifier entrance is 5 MW. Moving the monochromator to section U27 increases the seed signal to 150 MW. 
We note that in this case for 4 keV photons we reach saturation and generate 30 GW SASE signal, corresponding to 750 MW after the monochromator. We evaluated DBFEL performance under these conditions and found that it remains essentially unchanged with respect to the
case considered in the previous section.
\begin{figure}
\label{power-8kev}
\caption{Amplifier undulator peak power output at 8 keV as a function of distance $z$ compared to different seed power signals: 10 kW corresponds to the SASE case, 5 MW is the seed power with four crystal monochromator at U24 and 150 MW for the four crystal monochromator at U27. Yellow arrow indicates the end of the HXR undulator in the latter case. Dashed line corresponds to the coherent power value given by Eq. \eqref{power-coherent} for the beam parameters provided in Tab. \ref{beam-parameters}.}\includegraphics[width=0.6\linewidth]{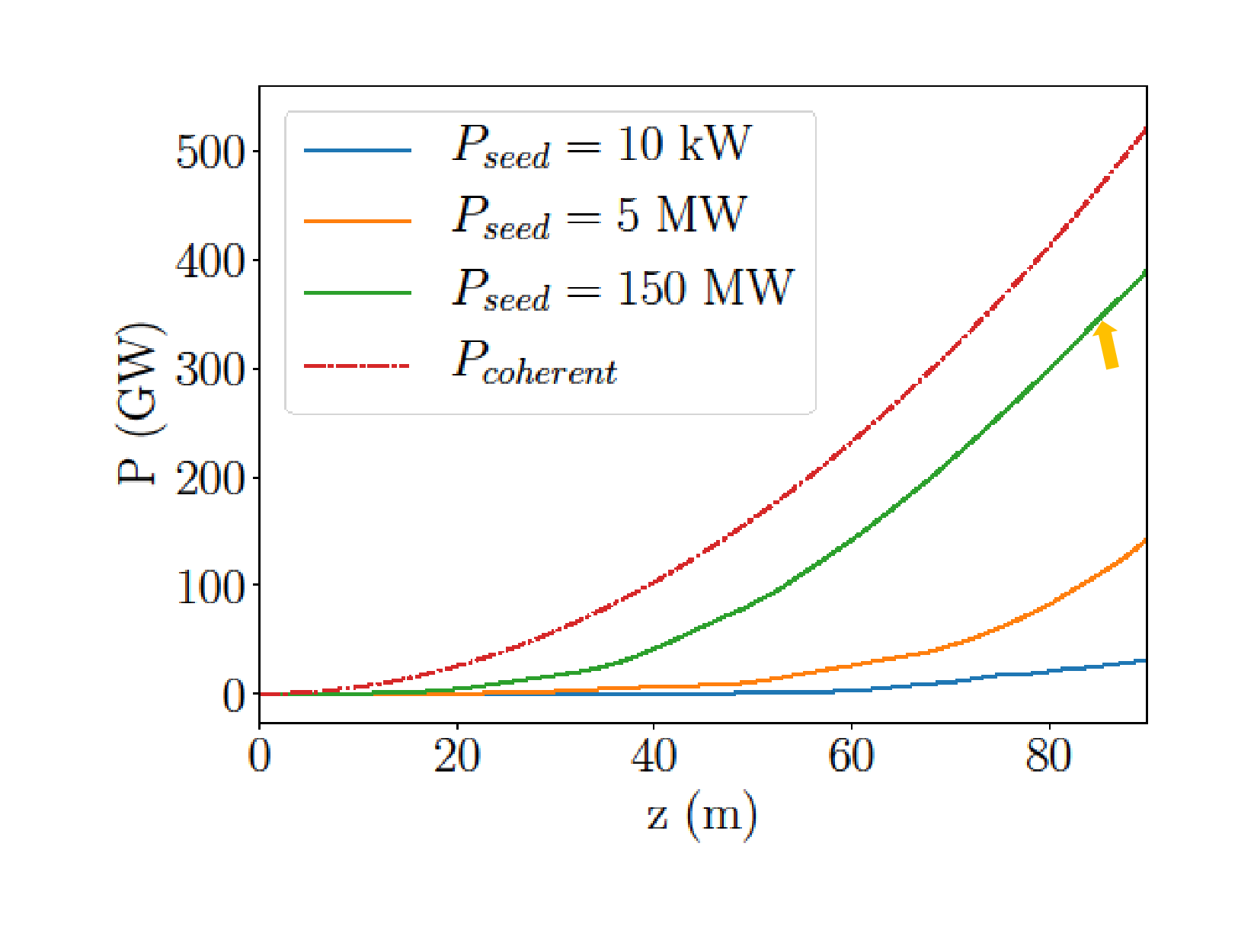}
\end{figure} 
The results of the simulations for the 8 keV case are shown in Figs. \ref{power-8kev}, \ref{8kev-harmonics}. 
For the 5 MW input seed case we obtain an output power of 135 GW, and for the 150 MW seed we get about 400 GW. When we account for the amplifier being three sections shorter, we obtain about 350 GW of power; see Fig. \ref{power-8kev}.
The spectral harmonics are displayed in Fig. \ref{8kev-harmonics}. For the case of a 150 MW input seed we have about 6 GW of power stored in the third harmonic at 24 keV, after reducing the amplifier length by three undulator sections. 
Finally, the power spectrum is presented in Fig. \ref{8kev-spectra}. 
The amount of power stored in the fundamental harmonic for the first case is 88\% and the latter case is 96\%. 

\begin{figure}\label{8kev-spectra}
 \caption{Power spectrum of 8 keV photons for the case of 5MW input seed (left panel) and 150MW input seed (right panel) for 8 keV photons. X-ray power profile in the time domain (bottom panel).}
 \includegraphics[width=0.45\linewidth]{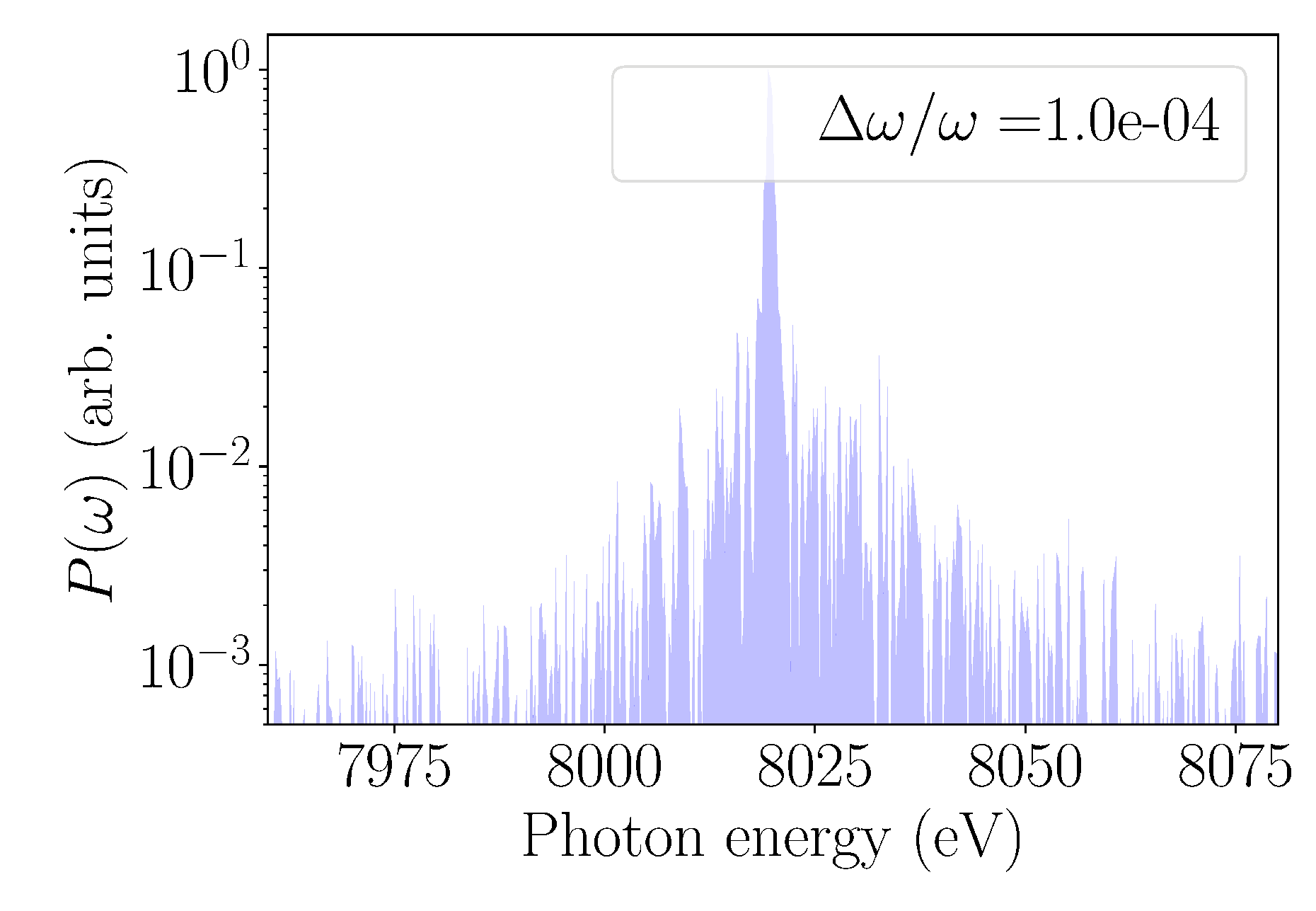}
  \includegraphics[width=0.45\linewidth]{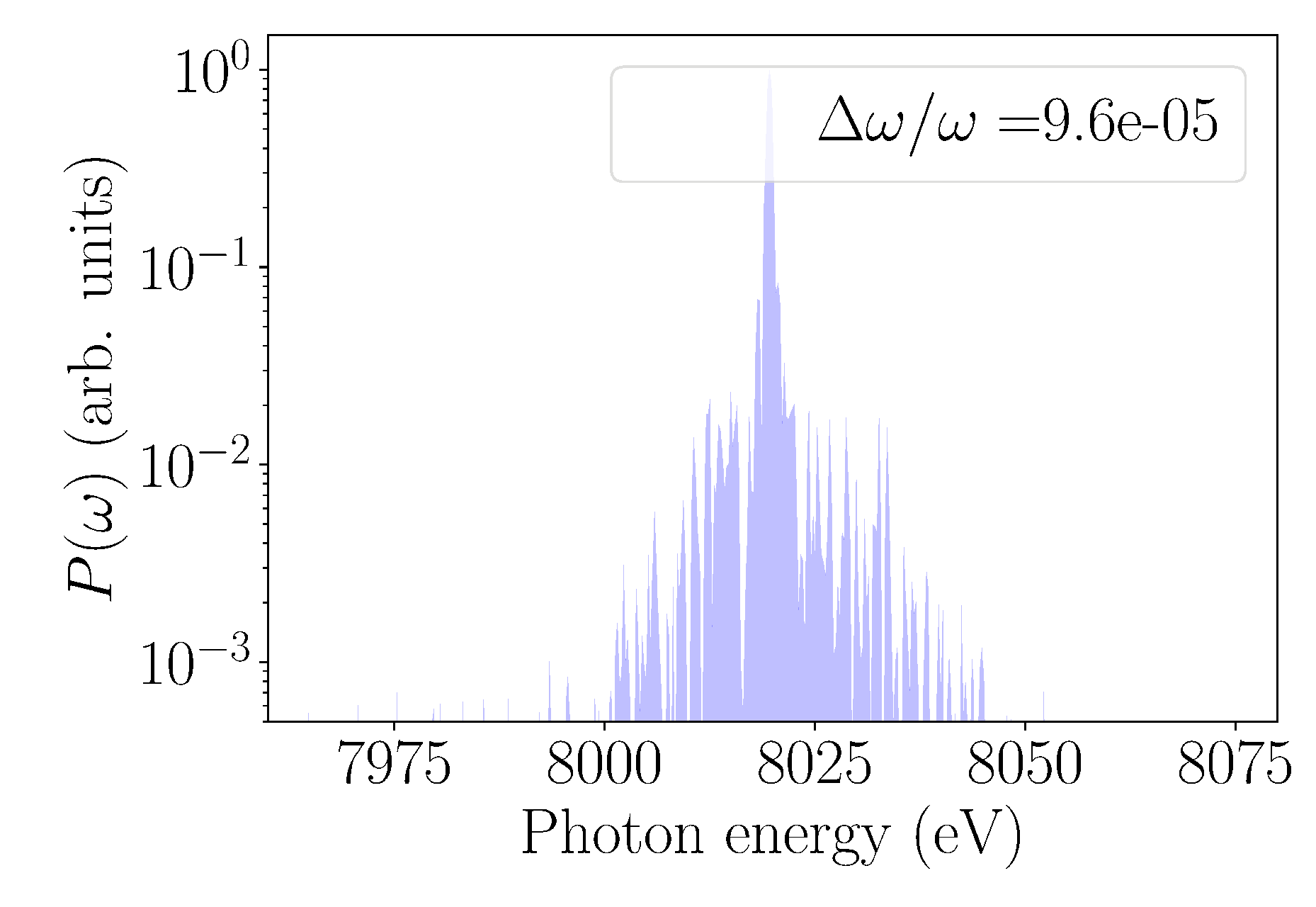}\\\includegraphics[width=0.5\linewidth]{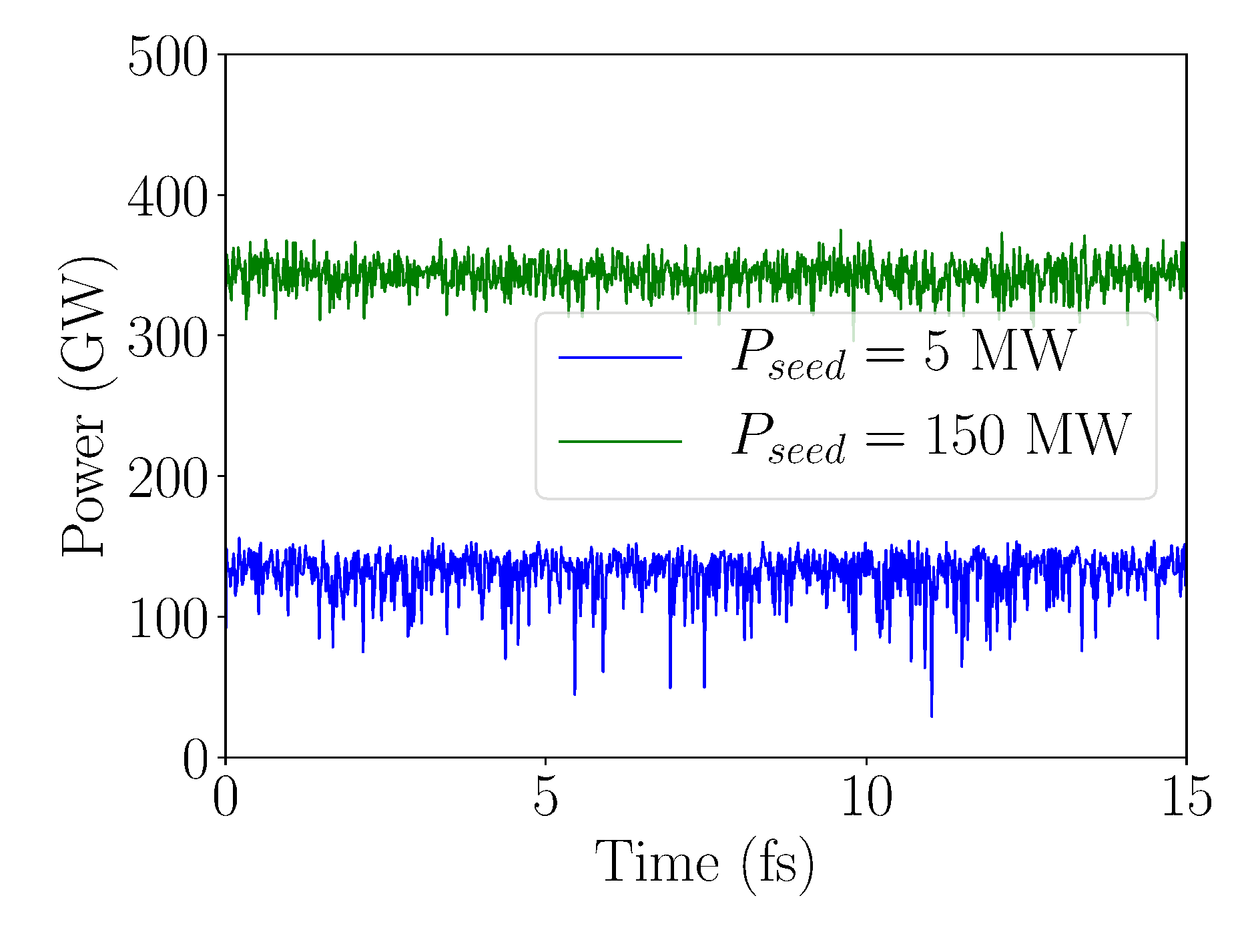}

\end{figure}

\begin{figure}\label{8kev-harmonics}
 \caption{Third and fifth harmonic of 8 keV photons power as a function of distance $z$ in the amplifier undulator for 5MW (left panel) and 150 MW (right panel) input seed power.}
 \includegraphics[width=0.445\linewidth]{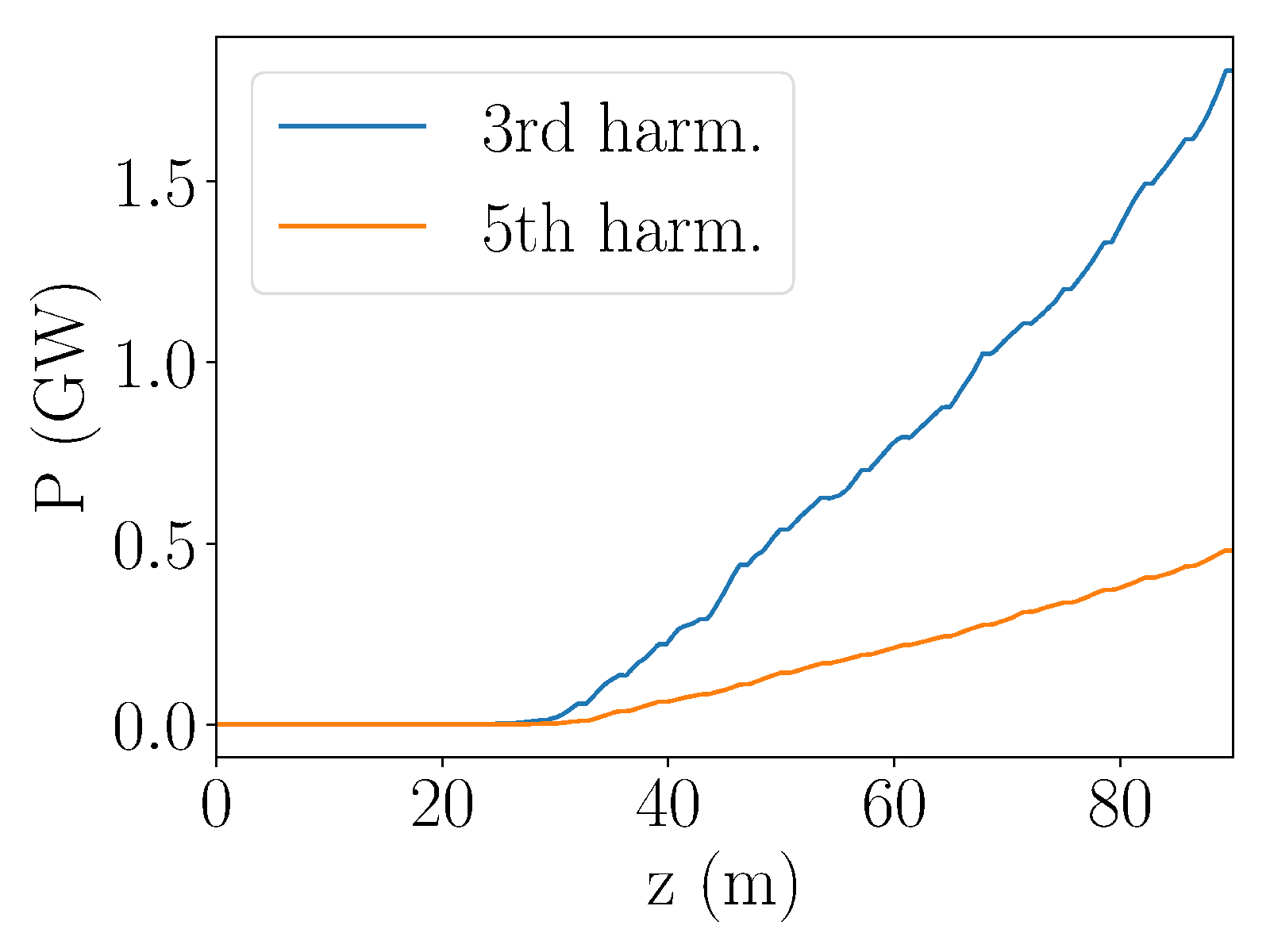}
  \includegraphics[width=0.445\linewidth]{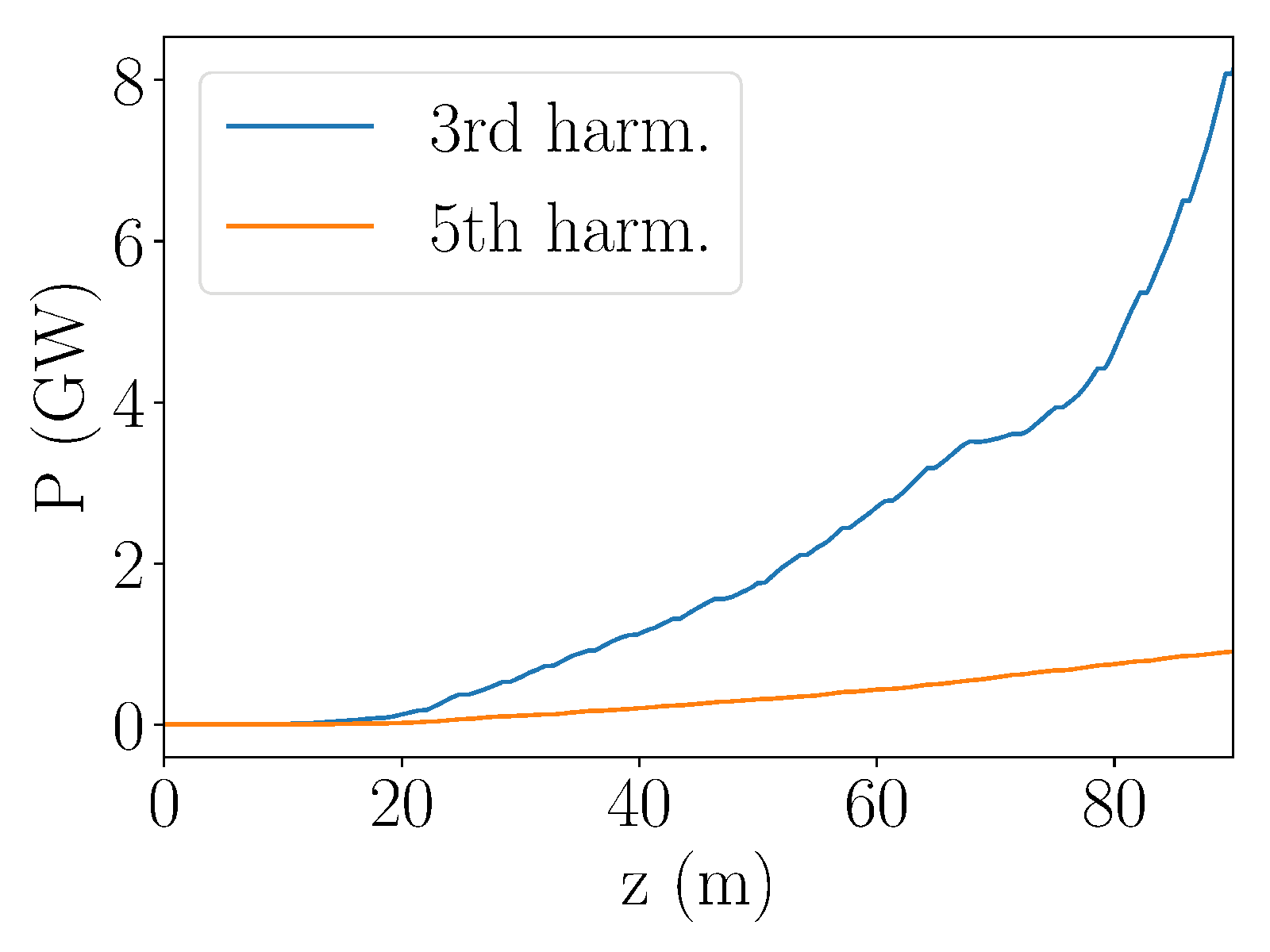}
\end{figure}

\begin{figure}
 \label{8kev-power-scan}
 \includegraphics[width=0.54\linewidth]{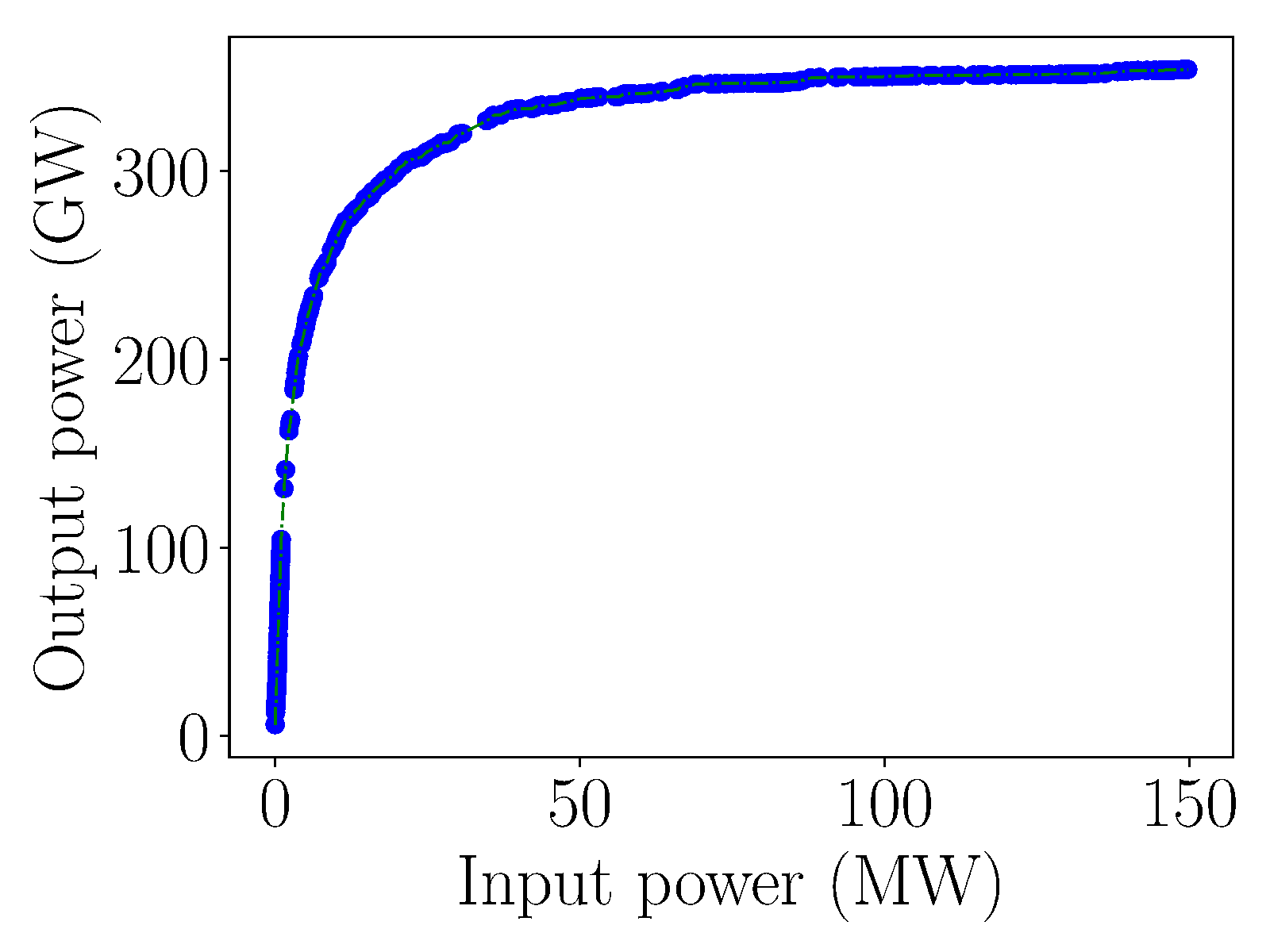}
 \caption{Output X-ray power as a function of input seed power for 8 keV photons.}
\end{figure}

For the 8keV case we have also evaluated the dependence of the output power on the seed power, as shown in
Fig. \ref{8kev-power-scan}. For the given beam parameters the output power starts to saturate at around 50 MW, corresponding to a peak SASE
power of 2 GW, obtainable by moving the four crystal monochromator by only two undulator sections.

\subsection{Comparison with double slice self-seeding: energy spread, emittance effects}
\label{fresh-slice}
\begin{figure}
\label{beam-scan}
\caption{Output power as a function of transverse beam emittance (left panel) and energy spread of the second bunch (right panel).}\includegraphics[width=0.425\linewidth]{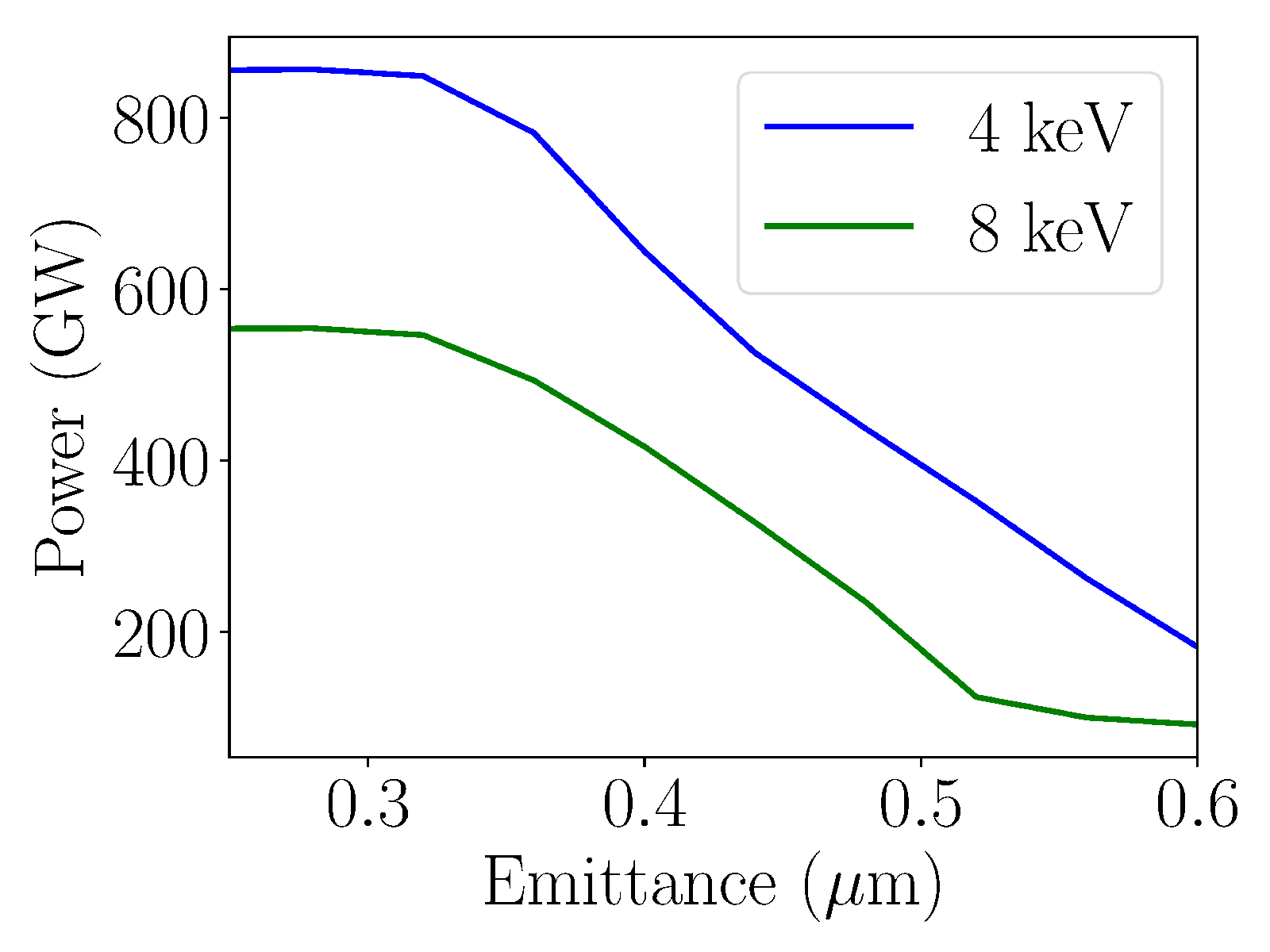}
\includegraphics[width=0.435\linewidth]{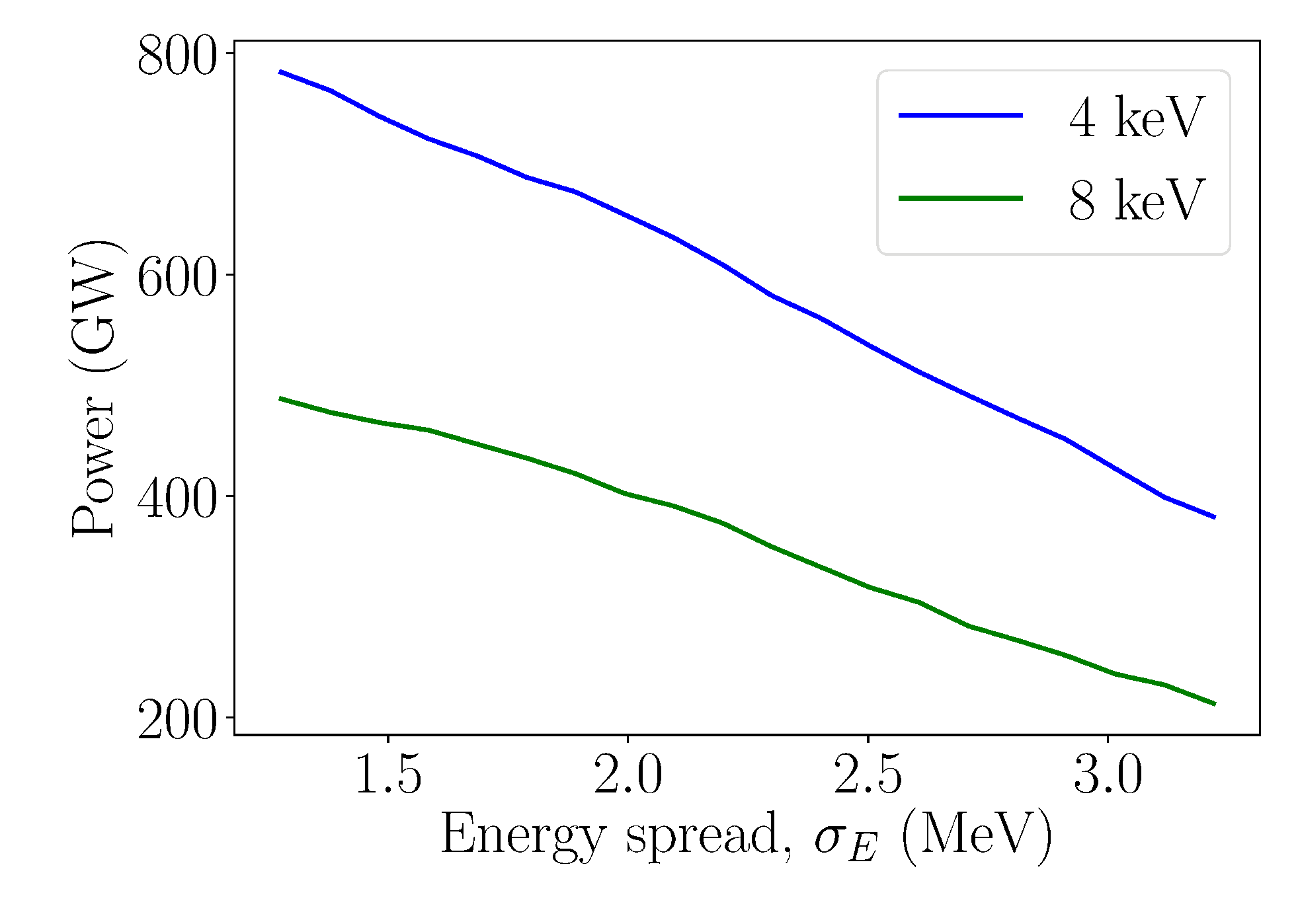}
\end{figure}

To evaluate the effects of the energy spread and emittance on the output power, we performed parametric scans for the cases of 4 keV and 8 keV photons; see Fig. \ref{beam-scan}.
The results, as expected, are strongly dependent on these two parameters. We notice that decreasing the energy spread to 1.5 MeV or less, the output power
becomes equal to the coherent power in Figs. \ref{power-4kev}, \ref{power-8kev} proving this parameter to be of critical importance in determining the DBFEL performance.
Figure \ref{beam-scan} also provides a comparison of the proposed DBFEL scheme with the existing double slice single bunch FEL \cite{EmmaC}, which already carries the brightness increase over the single bunch case.
In the double slice configuration, only about 1/3 of the bunch is used to generate the SASE signal and another 1/3 for the amplification process.
The remaining 1/3 of the bunch mostly contributes to the spectral background by increasing the overall beam emittance and the energy spread of the lasing slice \cite{Craievich}. To compare this case with DBFEL we must triple the charge from 60 pC to 180 pC, thus increasing the beam emittance from 0.4 $\mu$m to 0.6 $\mu$m \cite{Ding:2009zzc}.
It can be seen from the Fig. \ref{beam-scan} that such an increase significantly lowers the X-ray output power, with respect to the DBFEL. 
Thus, with a minor change in the HXR beamline at higher photon energies, DBFEL far exceeds the double slice FEL scheme.
Alternatively,
any possible enhancement in beam quality leads to even better performance of the DBFEL.

\subsection{Shot-to-shot power fluctuations}
\begin{figure}
\label{shot-to-shot}
\caption{Shot-to-shot power fluctuations of 8 keV fundamental energy photons due to SASE seed signal (left) and histogram of power fluctuations (right).}\includegraphics[width=0.41\linewidth]{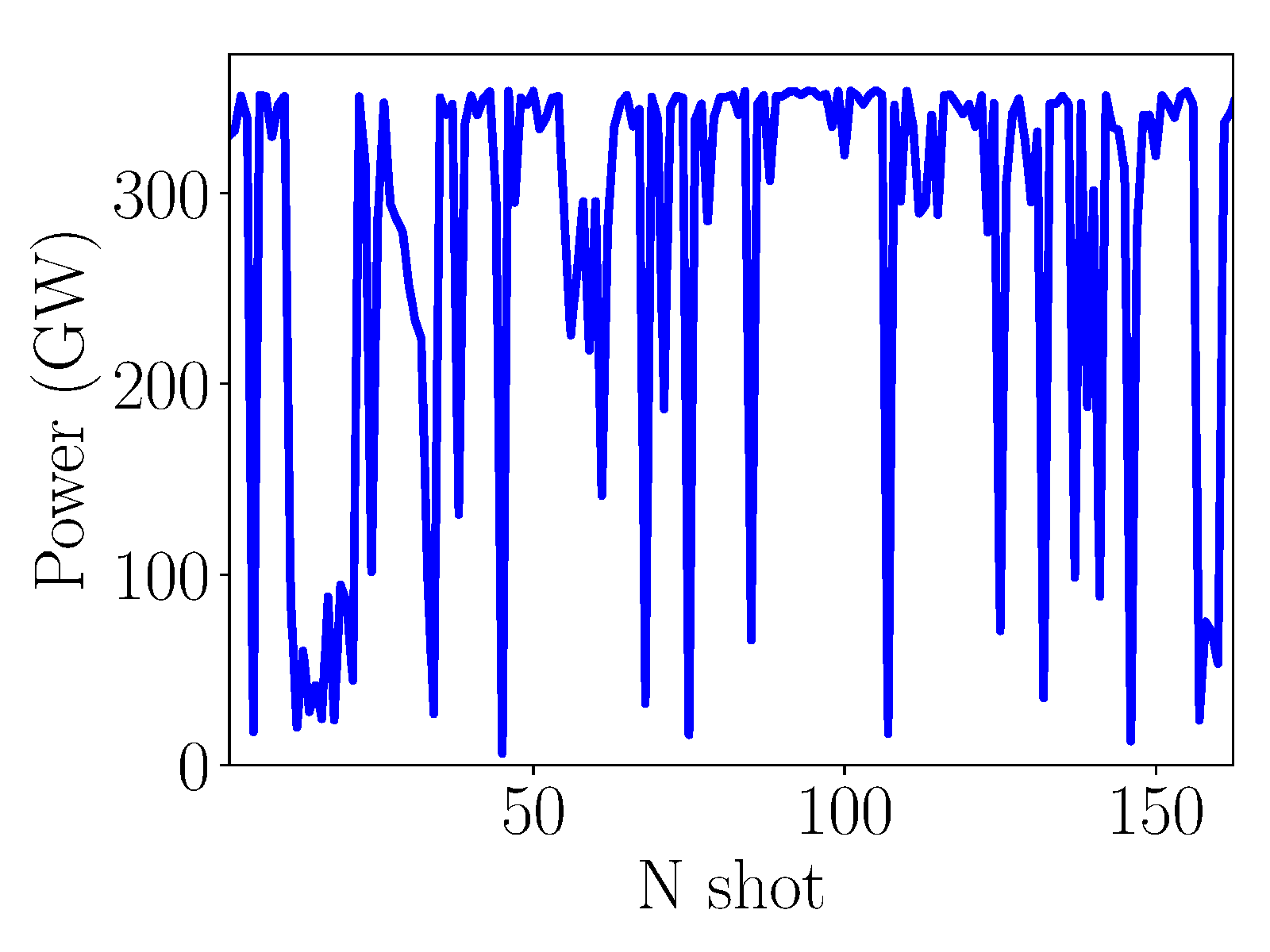}
\includegraphics[width=0.40\linewidth]{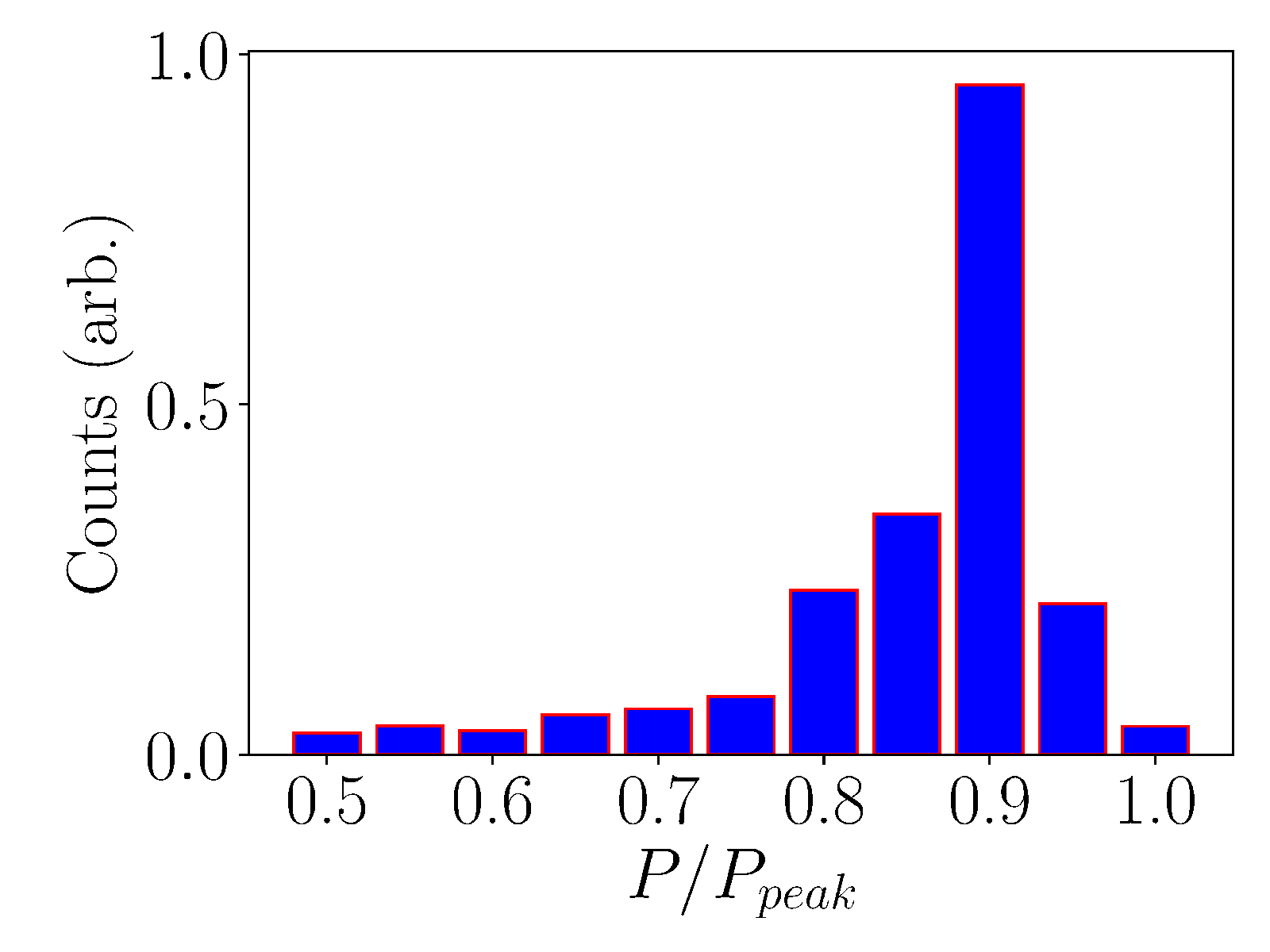}
\end{figure}

To estimate shot-to-shot power fluctuations of 8 keV X-rays in our DBFEL setup we used the spectrum provided in Fig. \ref{sase-sim-8kev} and the diamond (1,1,1) reflectivity curve shown in Fig. \ref{8kev-reflection}.
To convert the reflectivity curve into frequency domain we utilized the following relation, similar to \cite{Sun:yi5051}:
\begin{equation}
 \Delta \omega = -\omega ctg \theta_B \Delta\theta,
\end{equation}
where $\theta_B$ is the Bragg angle.
For the cases of 4 - 8 keV photons we found the reflectivity window width to be similar to the single SASE spike width. Thus, we also note that the input seed signal can be assumed to be pseudo-Gaussian in time.
To perform our calculations, we convoluted the SASE spectra and the crystal reflectivity curve directly for multiple realizations of SASE. The simulation results are presented in Fig. \ref{shot-to-shot}. As an effect of the tapering, the amplifier section is a very high gain system and saturates quickly, as displayed in Fig. \ref{8kev-power-scan}.
 One can notice the significant fluctuations of the resulting X-ray power, corresponding to the very narrow bandwidth of the monochromator crystals.
The methods to reduce these fluctuations will be the topic of our future studies. 

\subsection{AGU undulator}
\begin{figure}
\label{power-AGU}
\caption{AGU undulator peak power output at 8 keV as a function of distance $z$ compared to different seed power signals: 5 MW is the resulting four crystal monochromator seed power at U24 and 150 MW signal for the four crystal monochromator at U27. Dashed lined corresponds to the coherent power value given by Eq. \eqref{power-coherent} for the beam parameters provided in Tab. \ref{beam-parameters}.}\includegraphics[width=0.6\linewidth]{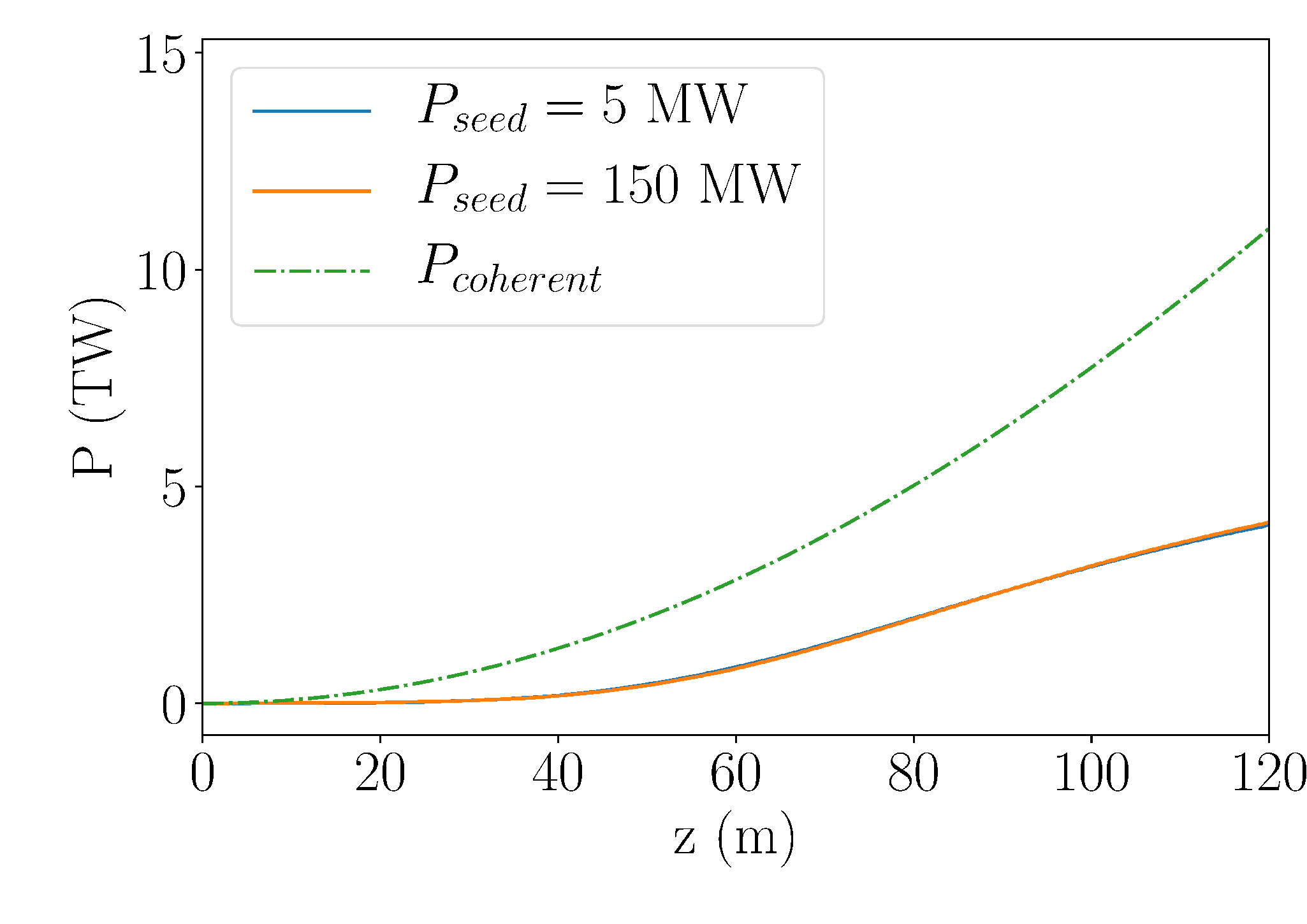}
\end{figure}

For comparison and to better understand the effects of the undulator design, we also consider the possible use of the Advanced Gradient Undulator (AGU) \cite{PhysRevAccelBeams.19.020705} as a second stage in our DBFEL system with the same beam parameters.
In brief, AGU is a proposed helical undulator based on a superconductor magnet technology and specifically designed for high X-ray power outputs. It is designed to have short drifts between undulator sections and provide strong electron beam focusing. We confine our studies to 8 keV fundamental photon energy. In this regime, we also consider two input seeds of 5 MW and 150 MW corresponding to the aforementioned cases of self-seeding. We confirm, via numerical simulations, that AGU, embedded in LCLS-II beamline, provides excellent X-ray output power in multi-TW range using the DBFEL scheme, even at the low input power level, as one can see in Fig. \ref{power-AGU}. 
Note that the resonant phase profile was similar to the one displayed in Fig. \ref{taper_profile}.

\section{Applications of tapered DBFEL}
\label{section-applications}
In this section we consider a few applications of the high power X-ray pulses generated in DBFEL. The applications are of course not limited to the ones discussed below. More applications will likely be developed once the system is in operation.

\subsection{Single particle imaging}
An X-ray pulse of 4 keV photons with 650 GW output power and 15 fs pulse duration contains about 10 mJ of energy. 
This value corresponds to about $1.5\cdot10^{13}$ coherent photons per pulse, a substantial increase with respect to what is achievable today and large enough for single particle
imaging \cite{Aquila}. At 8 keV, and assuming an output power of 50 GW or larger, this number is reduced by a factor of 3 to $5\cdot10^{12}$ coherent photons per pulse. We want
to remember that our assumption on the beam characteristics are rather conservative and
any operational improvement would lead to an even larger number of photons. It is also interesting to remark this number would be largely be increased in an AGU undulator. Lastly, we note that DBFEL can provide enough coherent photons for potential inelastic X-ray scattering experiments \cite{Chubar:yi5018}.

\subsection{Strong field electrodynamics}
\begin{figure}
\label{roadmap}
\caption{High power lasers. Power density and peak electric field. The star represents LCLS-II in DBFEL configuration
and photon beam focused to 10 nm spot size. Figure is a courtesy of P. Bucksbaum.}\includegraphics[width=0.52\linewidth]{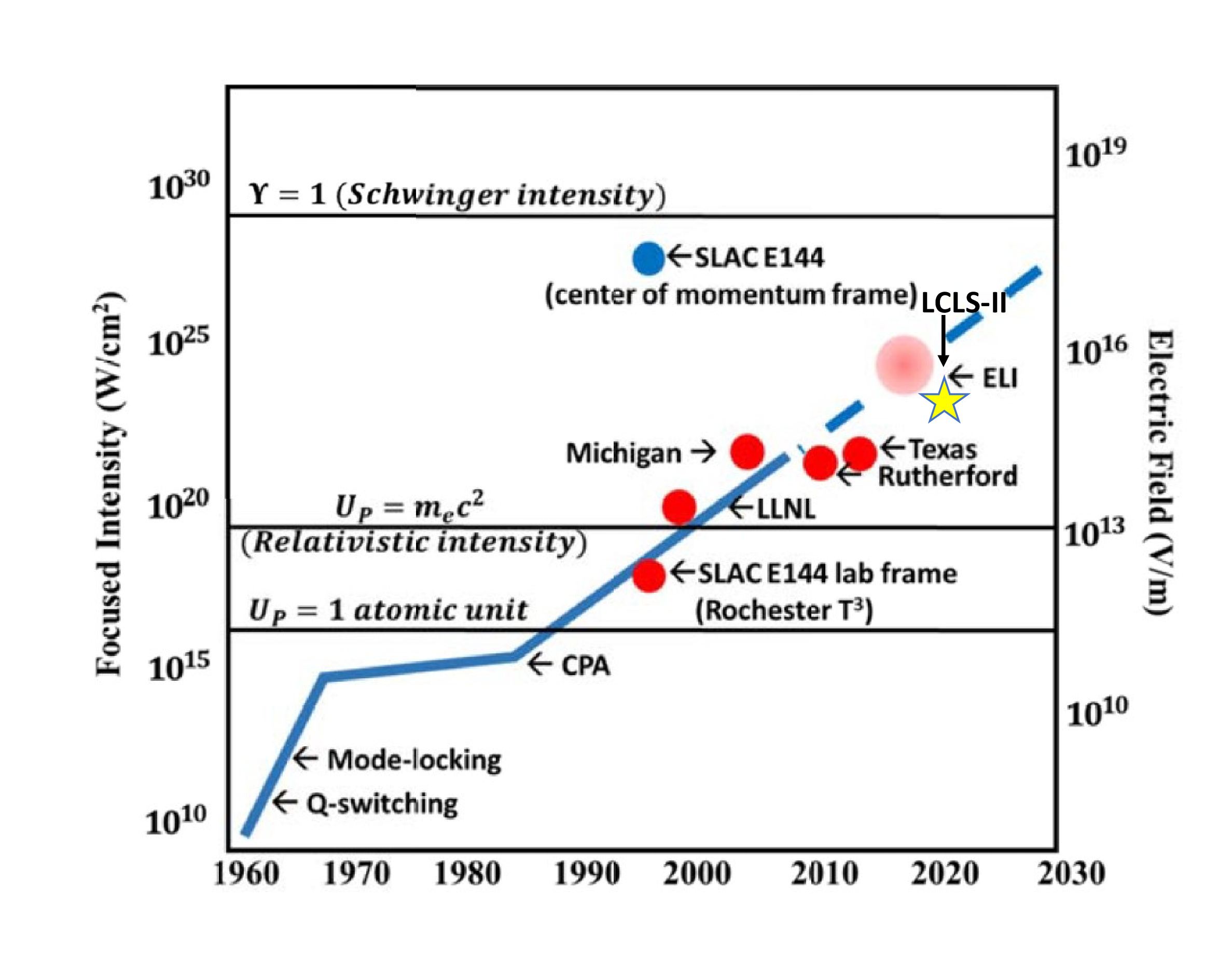}
\end{figure}

The development of very high power lasers at about 1 $\mu$m wavelength, reaching the PW power region, has opened new
capabilities for high field science. These opportunities have been recently reviewed in a National Academy of Science decadal
report \cite{NAP24939}. X-ray FELs can not reach the PW power level. However the X-ray pulse can be focused to a
much smaller spot size than the PW laser, tens of nm against few to ten $\mu$m, yielding similar power density
and peak electric field. 
The electric field gradient of $P_0=$ 1 TW X-rays focused to $\sigma_0=$10 nm spot is $E_0 = 1.2\cdot10^{15}$ V/m and the power density is $W_0 = 3.2\cdot10^{23}$ W/cm$^2$. 
The power density in W/cm$^2$ scales as
\begin{equation}
 W = \frac{P}{\pi \sigma^2} = 3.2\cdot10^{23}\frac{P/P_0}{(\sigma / \sigma_0)^2},
\end{equation}
while electric field gradient in V/m scales as
\begin{equation}
 E = \sqrt{\frac{PZ_0}{\pi\sigma^2}} = 1.2\cdot10^{15}\frac{\sqrt{P/P_0}}{(\sigma / \sigma_0)},
\end{equation}
where $P_0=1$ TW and $\sigma_0$=10 nm. We may view these numbers as reference and estimate the peak parameters of the tapered DBFEL. For the maximum of 650 GW of 4 keV photon peak power focused to 100 nm spot size, typical value presently obtainable, one gets $2.1\cdot10^{21}$ W/cm$^2$ of power density and $9.6\cdot10^{13}$ V/m field gradient. 
If possibly focused to a 10 nm spot size, a value recently achieved in a delicate state-of-the-art experiment at the XFEL
SACLA facility, in Japan \cite{2010NatPh, Yamauchi_2011}, the 4 keV pulse obtained in a DBFEL gives a power density of $2\cdot10^{23}$ W/cm$^2$ and a peak electric field of
$10^{15}$ V/m, values similar to those obtainable in a PW laser, as shown in Fig. \ref{roadmap} and in \cite{SFQEDslides}.
One can also consider backscattered HXR pulses that are additionally focused and collided head on with the electron beam, as was done in the E144 experiment at SLAC \cite{PhysRevLett.79.1626}.
In the electron rest frame the X-ray field gradient is multiplied by $\gamma$ and the power density by $\gamma^2$, yielding for a 6 GeV electron beam and 100 $\mu$m X-ray spot size
$W' = 3\cdot10^{29}$ W/cm$^2$ and $E' = 1.2\cdot10^{18}$ V/m.
If one recalls Schwinger critical field gradient $E_{cr} = \alpha m c^2/ e r_e \approx 1.3\cdot10^{18}$ V/m and $W_{cr}\approx 4 \cdot 10^{29}$ W/cm$^2$, a DBFEL generated X-ray signal backscattered with the electron beam can reach the regime where $E' / E_{cr}\equiv \chi \approx 1$. In addition, with
an improvement in X-ray focusing to 10 nm spot size, if possible in TW regime, one can reach $W' = 3\cdot10^{31}$ W/cm$^2$, $E' = 1.2\cdot10^{19}$ V/m, and $\chi\approx10$, presenting an opportunity to probe perturbative and non-perturbative strong-field QED effects, currently unavailable at modern XFEL facilities. We note that for PW lasers the normalized vector potential $a_0$ is an order of 1, while for X-rays it is smaller than 1, opening new and complimentary areas of exploration \cite{Ritus1985,RevModPhys.84.1177,PhysRevLett.110.070402}. 
Hence, LCLS-II offers the possibility of exploring at X-ray
wavelength most of the science that can be done with PW
lasers, like laser-plasma interaction, high energy density
science, planetary physics and astrophysics, and QED at
extreme fields above the Schwinger limit.

\section{Conclusions}
In conclusion, the presented DBFEL setup provides significant advantages over single bunch and fresh slice self-seeding schemes. We have demonstrated, via numerical simulations, that DBFEL can provide sub-TW X-ray pulses in the range of 4 keV to 8 keV with nearly transform-limited spectrum bandwidth. Improvements in the beam quality and increase in the peak current make it possible to reach near 1 TW peak power level, which enables many new high-field physics experiments. In addition, the proposed four crystal monochromator setup will benefit as well to the nominal single bunch self-seeding LCLS-II operations. 

     % Appendices appear after the main body of the text. They are prefixed by
     % a single \appendix declaration, and are then structured just like the
     % body text.

%\appendix
%\section{Appendix title}
%\subsection{Title}

     %-------------------------------------------------------------------------
     % The back matter of the paper - acknowledgements and references
     %-------------------------------------------------------------------------

     % Acknowledgements come after the appendices

\ack{Acknowledgments}\\
This work was supported by the U.S. Department of Energy Contract No. DE-AC02-76SF00515.
The authors are grateful to Yiping Feng, Yuantao Ding, Heinz-Dieter Nuhn, Juhao Wu, Zhirong Huang, Gennady Stupakov, Joe Duris, Gabriel Marcus, Chris Mayes, David Reis (SLAC) and Sebastian Meuren (Princeton University) for very useful and instructive discussions.

     % References are at the end of the document, between \begin{references}
     % and \end{references} tags. Each reference is in a \reference entry.
%\bibliographystyle{iucr}
%\bibliography{iucr}

%\begin{references}
%\reference{Author, A. \& Author, B. (1984). \emph{Journal} \textbf{Vol}, 
%first page--last page.}
%\end{references}

%\referencelist

     %-------------------------------------------------------------------------
     % TABLES AND FIGURES SHOULD BE INSERTED AFTER THE MAIN BODY OF THE TEXT
     %-------------------------------------------------------------------------

     % Simple tables should use the tabular environment according to this
     % model

     % Postscript figures can be included with multiple figure blocks

\end{document}